\documentstyle[]{mn}
\title[ Emission lines and optical continuum in low-luminosity 
radio galaxies]
{ Emission lines and optical continuum in low-luminosity 
radio galaxies}

\author[Wills et al.]
       {K.A. Wills$^{1}$, R. Morganti$^{2}$, C.N. Tadhunter$^{1}$,
T.G.Robinson$^{1}$, M. Villar-Martin$^{3}$ 
        \\
$^{1}$ Department of Physics and Astronomy, University of Sheffield, Hounsfield Road, Sheffield, S3 7RH, UK\\
$^{2}$ Netherlands Foundation for Research in Astronomy, Postbus 2, 7990 AA Dwingeloo, The
Netherlands \\
$^{3}$ Department of Physical Sciences, University of Hertfordshire, College
Lane, Hatfield, Herts AL10 9AB, UK }

\date{}
\def\ltsim{\ifmmode\stackrel{<}{_{\sim}}\else$\stackrel{<}{_{\sim}}$\fi}
\def\gtsim{\ifmmode\stackrel{>}{_{\sim}}\else$\stackrel{>}{_{\sim}}$\fi}

\begin{document}
\maketitle
\begin{abstract}

We present spectroscopic observations of a complete sub-sample of 13
low-luminosity radio galaxies selected from the 2Jy sample. The
underlying continuum in these sources is carefully modelled in order
to make a much-needed comparison between the emission line and
continuum properties of FRIs with those of other classes of radio
sources. We find that 5 galaxies in the sample show a measurable UV
excess: 2 of the these sources are BL~Lacs and in the remaining 3
galaxies we argue that the most likely contributor to the UV excess is
a young stellar component.  Excluding the BL~Lacs, we therefore find
that $\sim$ 30\% of the sample show evidence for young stars, which is
similar to the results obtained for higher luminosity samples. We
compare our results with far-infrared measurements in order to
investigate the far-infrared-starburst link. The nature of the
optical-radio correlations is investigated in light of this new
available data and, in contrast to previous studies, we find that the
FRI sources follow the correlations with a similar slope to that found
for the FRIIs. Finally, we compare the luminosity of the emission
lines in the FRI and BL~Lac sources and find a significant difference
in the [OIII] line luminosities of the two groups. Our results are
discussed in the context of the unified schemes.

\end{abstract}


\section{Introduction}
 
Following the work of Fanaroff \& Riley (1974), it is now well
understood that extended radio galaxies appear to come, broadly
speaking, in two different flavours: edge-darkened, low radio power
and powerful ($>$ 2.5 $\times$ 10$^{26}$ W Hz$^{-1}$ at 178 MHz; Baum,
Zirbel \& O'Dea 1995), edge-brightened. The former are usually know as
Fanaroff-Riley type I (FRI) and the latter as Fanaroff-Riley type II
(FRII). However, the nature of this dichotomy remains unclear: whether
this is related to the nature of the nuclear engine (perhaps the
black-hole mass or spin), the mass accretion rate, or even on a larger
scale the conditions of the ISM.

Resolving this issue requires knowledge of the characteristics of the
emission (in different wavebands) from both types of radio galaxies;
understanding the mechanisms that may produce the emission and explaining
the differences.  A number of studies have recently concentrated on
understanding the nuclear characteristics of FRIs using optical data. From
HST observations, the presence of an optical non-thermal component has
been confirmed by the detection of central compact cores (CCCs) in HST
images of FRIs (Chiaberge, Capetti \& Celotti 1999). The correlation
between the CCC's optical flux and the flux of the radio core for
FRI/BL~Lacs objects (Chiaberge et al. 1999) indicates that these cores
are due to optical synchrotron radiation produced in the inner region of a
relativistic jet.  This is therefore consistent with the idea that FRIs
are the parent population of the BL~Lacs, as predicted by the unified
schemes (Urry \& Padovani 1995).  Moreover, the detection of optical cores
in a large fraction of FRI radio galaxies suggests that the circumnuclear
disks, if present, must be geometrically thin, unlike the optically and
geometrically thick tori which are an essential ingredient of the unified
schemes for powerful FRIIs/radio-loud quasars. This already points to
some, possibly intrinsic, difference in the nuclear regions of FRI and
FRII radio galaxies.

Another powerful indicator of the activity in the nucleus is the warm
ionised gas in the circumnuclear regions.  This can be studied using the
optical emission lines, and indeed has been extensively used in FRII to
test the unified schemes hypothesis and to understand the ionization
mechanism for the gas.  Partly as a result of the low luminosity of their
emission lines, only sparse spectroscopic data are available for the FRIs.
Compared to FRIIs, FRIs have (on average) 5 to 30 times weaker emission
lines (for a similar radio total luminosity, Baum et al. 1995). This has
therefore limited the study of the characteristics of these galaxies.  
Indeed, FRIs are invariably classified as weak-line radio galaxies (WLRG).  
A tight correlation between radio core emission and H$\alpha$+[NII]
emission in FRIs has recently been confirmed by Verdoes Kleijn et al.
(2002) using HST narrow band images. This has been interpreted as a strong
indication that the emission gas is excited by an AGN-related process and
not only due to processes associated with the host galaxy (e.g. old
stars).

An interesting `complication' is the suggestion that FRIs and FRIIs follow
separate correlations between optical line luminosity and radio luminosity
(Baum et al. 1995). However, from this published data it is not clear the
extent to which this is really the case and in a later paper, Tadhunter et
al. (1998) argue that only the [OIII] correlation, and not the [OII]
correlation, indicates a difference between FRI and FRII sources. It is
important to point out, however, that the Tadhunter et al. study, although
based on a well-defined sample, resulted in many upper limits in the data.
Part of the problem with previous studies has been the lack of accurate
continuum subtraction, which is essential for measuring accurate fluxes
and upper limits for the faint emission lines present in the spectra of
FRI sources.  Therefore, in order to advance our understanding of the FRI
sources further, it is important to make higher quality observations of a
well-defined sample and also to model and subtract the underlying
continuum emission.

The analysis of the optical continuum is itself an important component
in understanding the characteristics of FRIs relative to FRIIs.  Much
progress has already been made in analysing the optical continua of
powerful FRII radio galaxies, particularly in understanding the nature
of the UV excess that is often present in these galaxies (Tadhunter,
Dickson \& Shaw 1996, Robinson et al. 2000, Aretxaga et al. 2001,
Wills et al.  2002, Tadhunter et al. 2002).  These studies suggest
that approximately 30\% of the observed host galaxies of FRIIs show a
significant contribution from a young stellar population component ---
probably related to starbursts induced in the merger events which
triggered the activity. However, no systematic studies have been done
for FRI sources, although at least one FRI is known to have a young
stellar component (Melnick, Gopal-Krishna \& Terlevich 1997, Aretxaga
et al. 2001), and a significant fraction of FRIs are also known to
exhibit UV continuum excess (e.g. Tadhunter et al. 2002). Clearly, in
order to investigate whether the triggering events are similar in the
two types, it is important to compare the optical/UV continuum
properties of the FRI and FRII host galaxies.  Again, such studies
have been hampered in the past by the poor quality of the optical data
for the FRI sources.

Here we present new, high quality optical spectra collected for a
complete sub-sample of 13 low-luminosity radio galaxies selected from
the Tadhunter et al. (1993, 1998) sample of 2Jy radio sources.  These
data increase substantially the number of FRIs for which good quality
spectra are available, and they allow us to compare the emission line
and continuum properties of the FRIs with those of other classes of
radio sources such as FRIIs and BL~Lacs.

A Hubble constant of H$_0$ = 50 km s$^{-1}$ Mpc$^{-1}$ and a
deceleration parameter of q$_0$ = 0 are assumed throughout.

\section{The observations}

The sample chosen for the purposes of this study comprises all objects
with FRI or core/halo radio morphologies, or classified as BL~Lac objects
from the complete sample of 2Jy radio sources described in Tadhunter et
al. (1993, 1998), with RAs in the range 04h$<$RA$<$17h (with the exception
of Cen A). Some general properties of this sample are listed in Table 1.
Most of the 14 sources in the sample fall at the lower end of the radio
power range for the 2Jy sample, and all of the objects are at relatively
low redshifts (z $<$ 0.06).  Spectroscopic observations for thirteen
galaxies (excluding 0521-36 because of a lack of time) were obtained with
the EFOSC2 at the ESO/MPG 2.2m telescope in February 1996.

\begin{table*}
\centering
\caption{Details of the radio galaxies in the sample.  Column 3 gives
the redshifts which are taken from either Tadhunter et al. 1993
($^{\dag}$), our own spectra ($^{\ddag}$) or the NASA/IPAC
Extragalactic Database (NED) ($^{\star}$), depending on which value
gives the best fit to the continuum (see also \S 3.4).  The 4.8 GHz
radio power (total and core) are taken from Morganti, Killeen \&
Tadhunter (1993) and the morphological classification comes from the
images in Morganti et al. (1993, 1999) (I = FRI, C+D = core + diffuse
emission, CJ = core/jet). VLBI images of some of the sources are also
presented in Venturi et al. (2000). The $R_{2.3 {\rm GHz}}$ parameter
(column 7) is defined as R = S$_{core}$/(S$_{tot}$-S$_{core}$) and
comes from Morganti et al. (1999). The E(B-V) values for galactic
reddening have been derived from the full-sky dust map presented in
Schlegel, Finkbeiner \& Davis (1998).}

\label{tab1}
\begin{tabular}{lccccccc}
\hline\hline\\ 
{\bf Name } & & {\bf z} & {\bf Radio} & ${\bf \log
P_{tot}}$ & ${\bf \log P_{core}}$ &  ${\bf R_{2.3 GHz}}$ & {\bf E(B-V)} \\ 
& & & {\bf Morphology} & {\bf W/Hz} & {\bf W/Hz} &  & {\bf mag.} \\
\hline    
0427-53 &         & 0.0412$^{\star}$  & I &   25.33& 23.56 & 0.0169 & 
0.012 \\
0453-20 &OF-289   & 0.035$^{\dag}$    & I &   24.99& 23.34 &  0.112 & 
0.041 \\ 
0521-36 &         & 0.055$^{\dag}$    & CJ &  26.10 & 25.28 & 0.176  & 0.038 \\
0620-52 &         & 0.051$^{\dag}$    & I &  25.17 & 24.49 & 0.062  & 
0.068 \\
0625-35 &OH-342   & 0.0525$^{\ddag}$  & I &  25.46 & 24.91 & 0.227  & 
0.067 \\
0625-53 &         & 0.0556$^{\ddag}$    & I &  25.40 & 23.75 & 
0.0078&0.094 \\
0915-11 &Hydra~A  & 0.0542$^{\ddag}$  & I &  26.26 & 24.46 & 0.0067  
&0.042 \\
1216+06 &3C~270   & 0.00747$^{\star}$ & I &  24.16 & 22.69 & 0.0194  
&0.018 \\
1246-41 &NGC4696  & 0.00994$^{\ddag}$ & C+D &  23.68 & 21.81 & 0.0068&0.113  \\
1251-12 &3C~278   & 0.0157$^{\ddag}$  & I &  24.39 & 22.94 & 0.016   
&0.053  \\
1318-43 &NGC~5090 & 0.0117$^{\ddag}$  & I &  23.97 & 23.49 & 0.153   
&0.144  \\
1333-33 &IC~4296  & 0.013$^{\dag}$    & I &  24.67 & 23.34 & 0.019   
&0.062 \\
1514+07 &3C~317   & 0.034$^{\ddag}$   & C+D &  24.73 & 24.34 & 0.163 &0.037  \\
1514-24 & ApLib   & 0.0480$^{\ddag}$  & CJ &  25.29 & 25.17 & 34     &0.138 \\
\hline \\
\end{tabular}
\end{table*}

The observations were carried out using a Thomson THX31156 1024 $\times$
1024 CCD with 19 $\mu$m (0.33$^{\prime\prime}$) pixels, a 2 arcsecond slit
and Grism \#3.  The Thomson chip has a gain of 2 e$^-$/ADU and the readout
noise is less than 5 e$^-$ rms. Use of Grism \#3 at an effective blaze of
3900 {\AA} resulted in a dispersion of 1.9 {\AA} pixel$^{-1}$ and an
approximate wavelength coverage of 3500-5400 {\AA}. The instrumental
resolution was $\sim$ 9.0 {\AA} with the 2 arcsecond slit. For each
source, the observations consisted of between two and four cycles of
typically 1200s per cycle (occasionally 300s, 600s or 900s cycles were
used). This resulted in a total of nearly 11 hours integration time for
the 13 radio galaxies. All but one of the sources were observed at
relatively low airmass (typically $<$1.1, but three sources 1.15-1.25).
Note that the source with the highest airmass (1514+07, 1.33) shows a
relatively poor fit in the red region of the spectrum, probably as a
result of differential refraction (see \S 3.4).

Reduction of the data proceeded along conventional lines using the
Starlink FIGARO package. This involved bias subtraction, cosmic ray
removal, flat-field correction, wavelength calibration, atmospheric
extinction correction, flux calibration and sky
subtraction. Flat-field correction was undertaken using dome flats
acquired with a 2 arcsec slit. Flux calibration was provided via
observations of the spectrophotometric standard stars EG 54, 76 and
248 using a 5.0 arcsec slit. Comparison between these different
standard star observations gives a relative flux calibration error of
$\pm$7 per cent across the wavelength range of each galaxy.

A typical extraction aperture of 2 arcsec (slit width) $\times$ 7.1
arcsec (21 pixels) was used in order to provide full coverage of the
nuclear regions of each source. The final rest-frame intensity spectra
are shown in Figure 1 (see Table 1 for redshifts used in the shift to
the rest-frame). The wavelength range of these spectra are equivalent
to the observed wavelength range 3500 - 5400 {\AA}, except in the case
of 0453-20 and 0620-52 where the observed wavelength range has been
cropped to 3700 - 5400 {\AA} due to light leakage on to the blue
region of the CCD. The spectra have been corrected for the effects of
galactic reddening using the Seaton (1979) extinction curve and the
values of E(B-V) derived from the full-sky dust map presented in
Schlegel, Finkbeiner \& Davis (1998, see Table 1). We have also
subtracted the nebular continuum (see Dickson et al. 1995) in the case
of 1514+07 and Hydra A, since these two are the only galaxies within
our sample to show a measurable H$\beta$ component.

In addition to our observations of the 13 radio galaxies, we also
present the results for the source 0521-36 (which we were unable to
observe due to time constraints) taken from Tadhunter et al. (1993)
and Boisson, Cayatte \& Sol (1989) for completeness.

\begin{table*}
\centering
\caption{Measurements of the 4000 {\AA} break (see \S 3.1), the flux
and luminosity of the emission lines (see \S 3.3) and the far-infrared
flux and luminosity (see \S 3.5) for the galaxies in the sample. In
the cases where the [OII] and [OIII] flux are both quoted as upper
limits, we do not calculate a [OIII]/[OII] ratio since its value is
not particularly meaningful. We quote the values for 0521-36 from
Tadhunter et al. (1993) where available. The far-infrared fluxes and
luminosities quoted have been calculated by co-adding scans from the
IRAS `all-sky' survey (see \S 3.5 for full details) except in the
cases (marked (P)) for which `pointed observations' are available
(Golombek, Miley \& Neugebauer 1988; Impey \& Neugebauer 1988). All
luminosities have been calculated assuming H$_0$ = 50 km s$^{-1}$
Mpc$^{-1}$ and q$_0$ = 0. }
\label{tab2} 
\begin{tabular}{lccccccccc}
\hline\hline\\ 
{\bf Name } & ${\bf D^{\prime}(4000)}$ & ${\bf \log F_{[OII]}}$
& ${\bf \log F_{[OIII]}}$ &
 ${\bf \log L_{[OII]}}$ &   ${\bf \log L_{[OIII]}}$ & {\bf [OIII]/[OII]} & {\bf F (60 $\mu$m)} & {\bf log L (60 $\mu$m)}   \\
& & ${\bf erg s^{-1}cm^{-2}}$ & ${\bf erg s^{-1}cm^{-2}}$ & ${\bf erg
s^{-1}}$ & ${\bf erg
s^{-1}}$ & & {\bf mJy} & {\bf W/Hz} \\
              &                 &          & & & &          &       \\
\hline    
0427-53 & 2.04 & $<$-14.98 & $<$-15.46  & $<$39.91 & $<$39.42 &  -   &  $<$ 48 & $<$ 23.5    \\
0453-20 & 1.82 & $<$-14.64 & $<$-15.31  & $<$40.10 & $<$39.43 &  -   &  690 & 24.5 \\
0521-36 &  -   & -14.49  & -13.99   & 40.66    & 41.16    & 3.13 &  354 (P) & 24.6 (P)  \\
0620-52 & 1.86 & $<$-14.71 & $<$-15.46  & $<$40.37 & $<$39.61 &  -   &  60 & 23.8   \\
0625-35 & 1.06 & -14.24 & -14.46    & 40.86    & 40.64    & 0.59 &  60 & 23.8  \\
0625-53 & 2.24 & $<$-14.80 & $<$-14.91  & $<$40.31 & $<$40.19 &  -   &  $<$ 63 & $<$ 23.9  \\
0915-11 & 1.43 & -13.79   & -14.42  & 41.34 & 40.71    & 0.24    &  155 (P) & 24.2 (P) \\
1216+06 & 2.01 & -14.12 & $<$-14.43   & 39.26   & $<$38.95  & $<$0.49 & $<$ 156 & $<$ 22.5  \\
1246-41 & 2.06 & -13.95   & -14.47  & 39.69  & 39.17   & 0.30       & 120 & 22.6 \\
1251-12 & 2.09 & -14.63   & -14.71  & 39.41  & 39.33   & 0.83       & $<$ 111 & $<$ 23.0 \\
1318-43 & 2.08 & -14.14   & $<$-14.60 & 39.64  & $<$39.18 & $<$0.35   & 160 & 22.9  \\
1333-33 & 2.09 & -14.09   & $<$-14.43 & 39.79 & $<$39.45 & $<$0.46    & 130 &  22.9  \\
1514+07 & 2.03 & -14.14   & -14.39  & 40.59 &  40.33   & 0.56       & $<$ 99 &  $<$ 23.6 \\
1514-24 & 1.18 & $<$-13.88 & -14.32 & $<$41.14  & 40.70 & $>$2.8    & 260 (P) & 24.4 (P) \\ 
\hline \\
\end{tabular}
\end{table*}

\begin{figure*}
\setlength{\unitlength}{1mm}
\label{fig1.fig}
\begin{picture}(10,225)
\put(0,0){\includegraphics{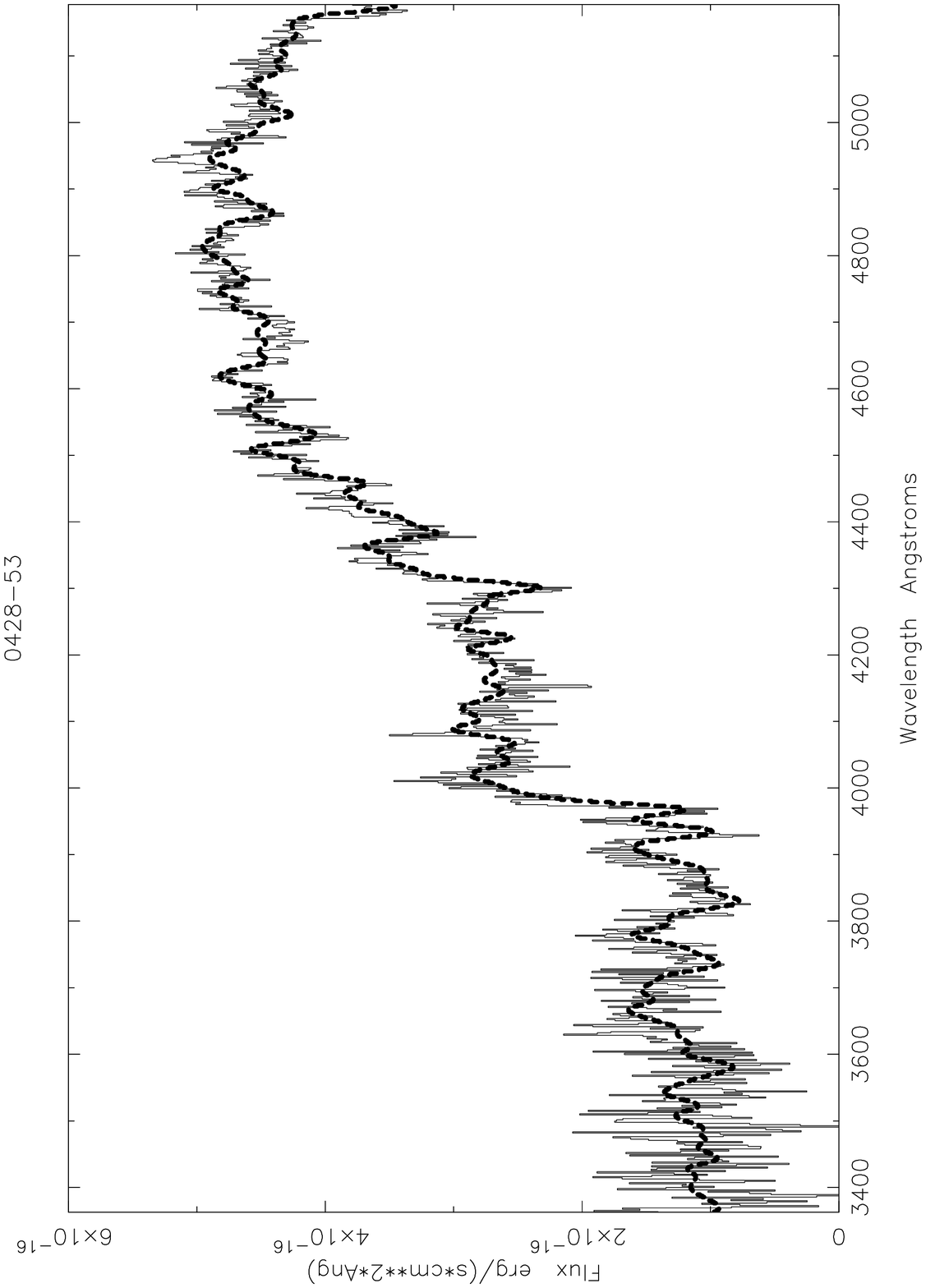}}
\put(0,0){\includegraphics{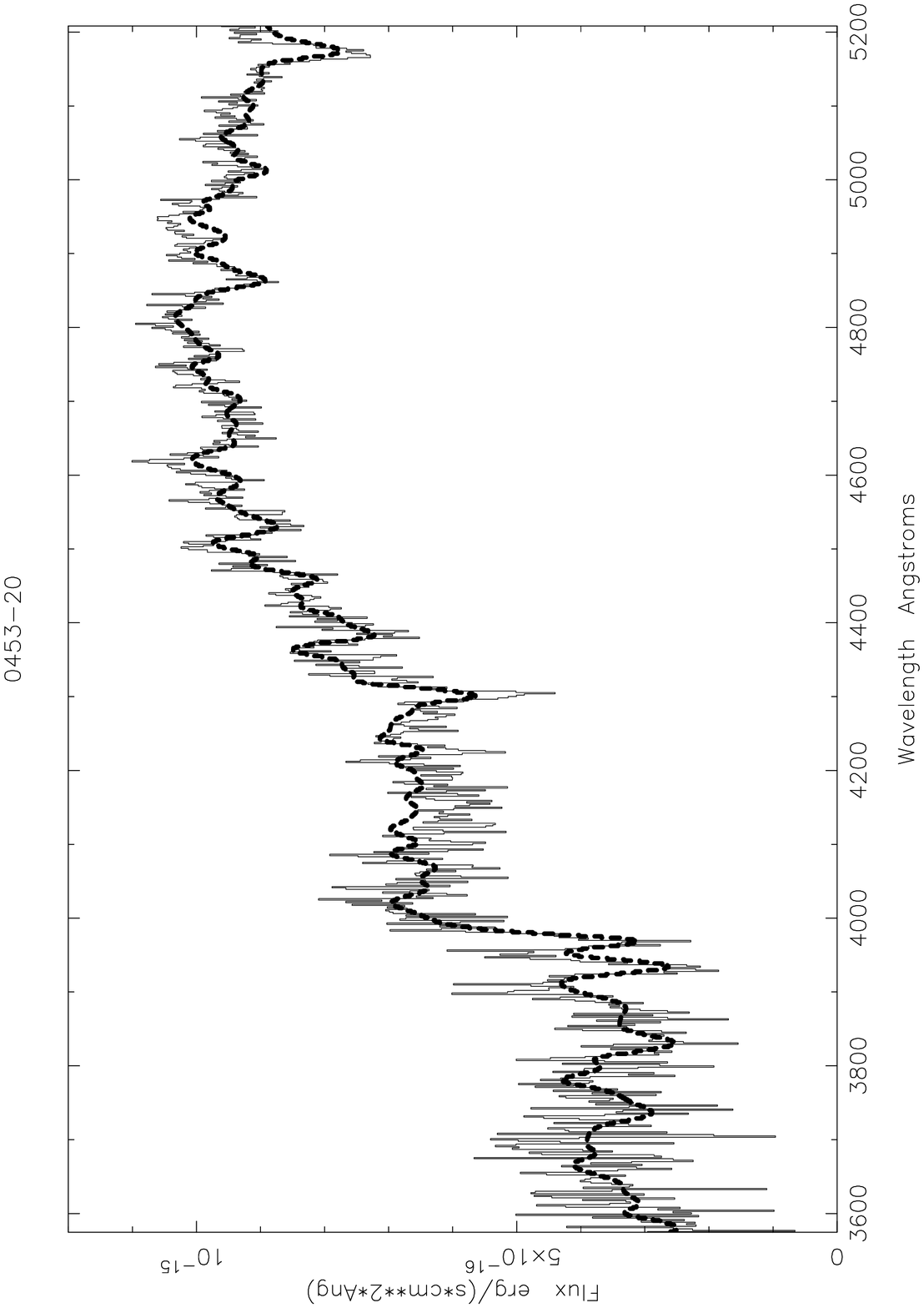}}
\put(0,0){\includegraphics{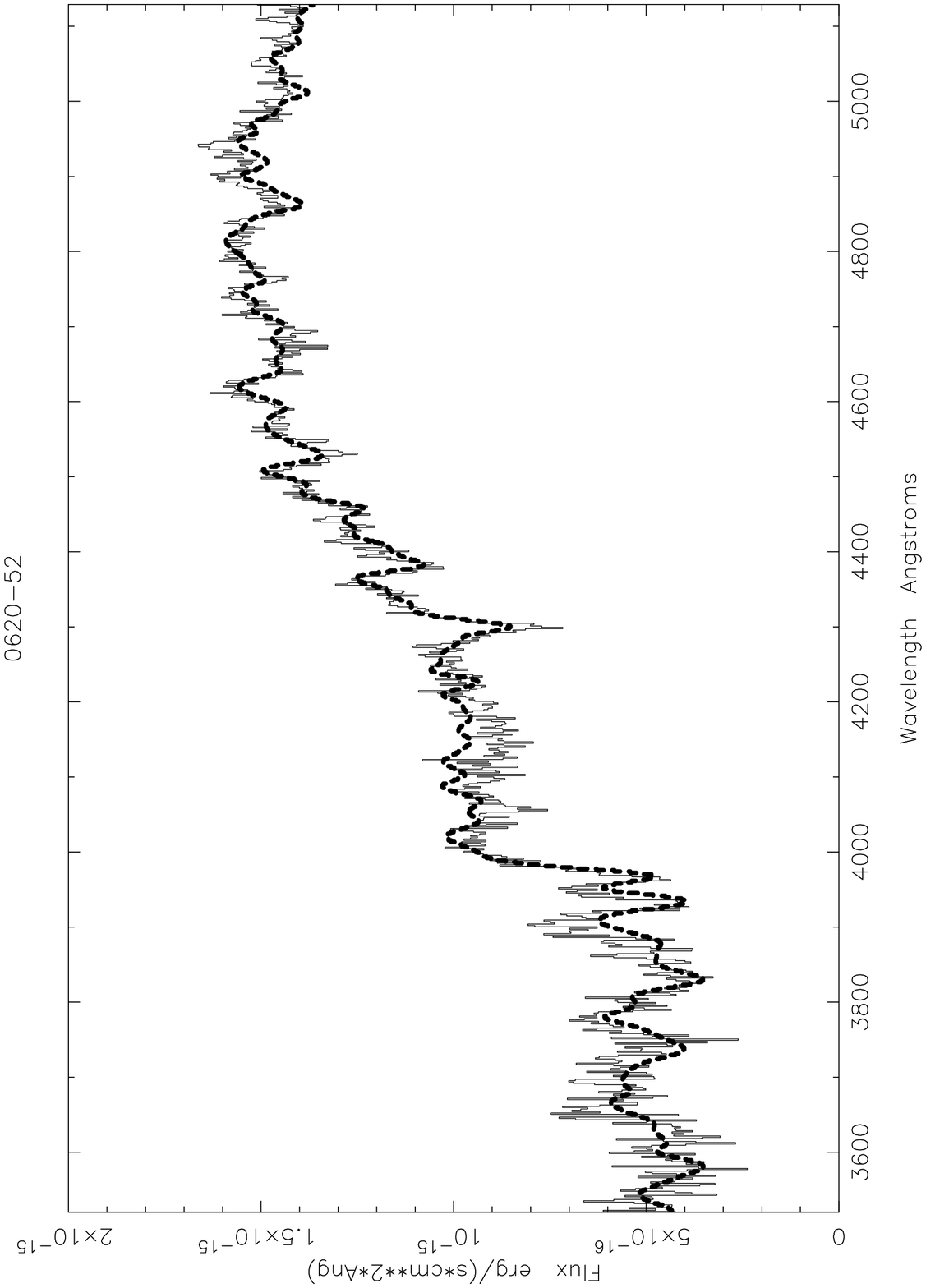}}
\put(0,0){\includegraphics{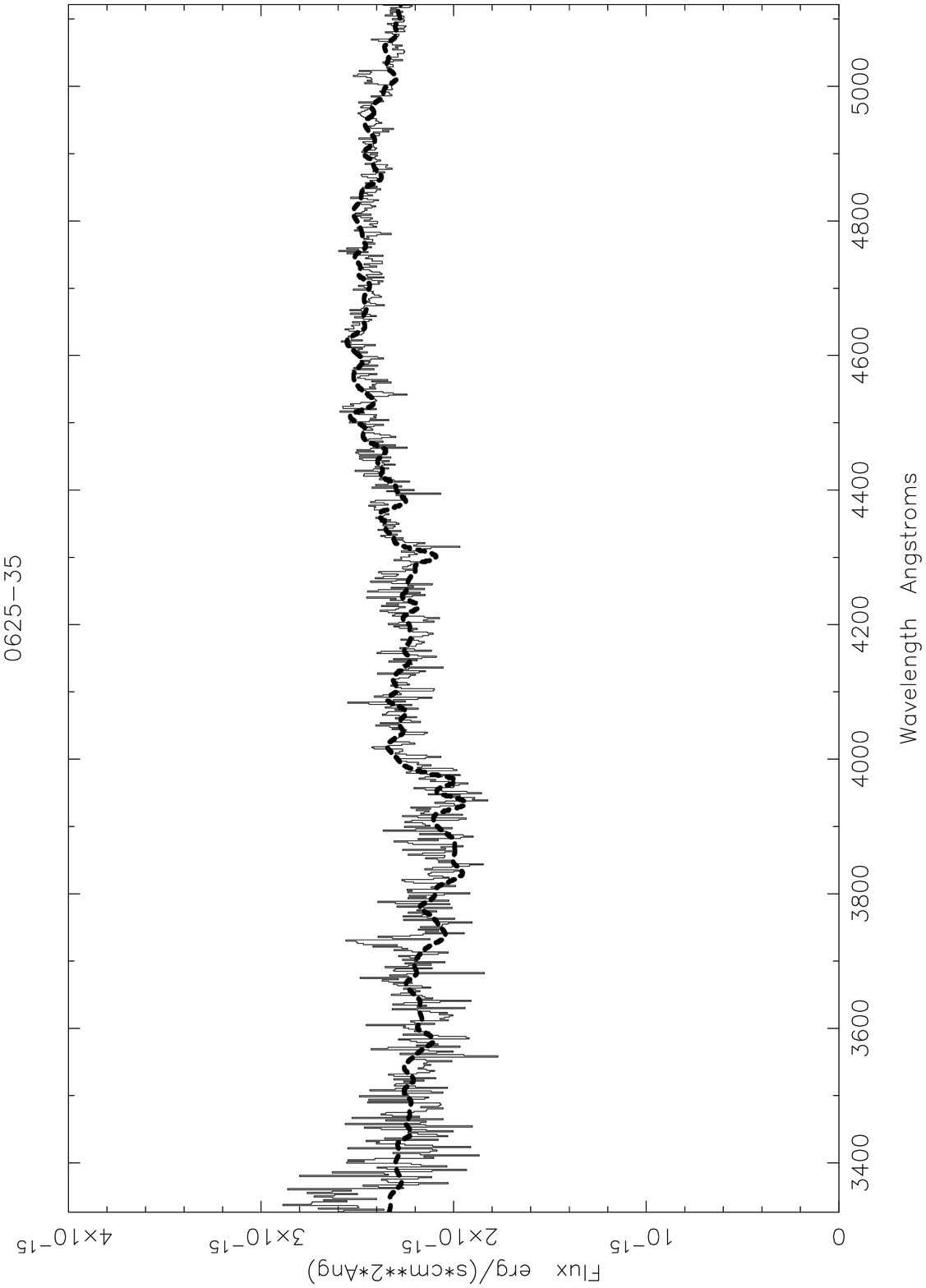}}
\put(0,0){\includegraphics{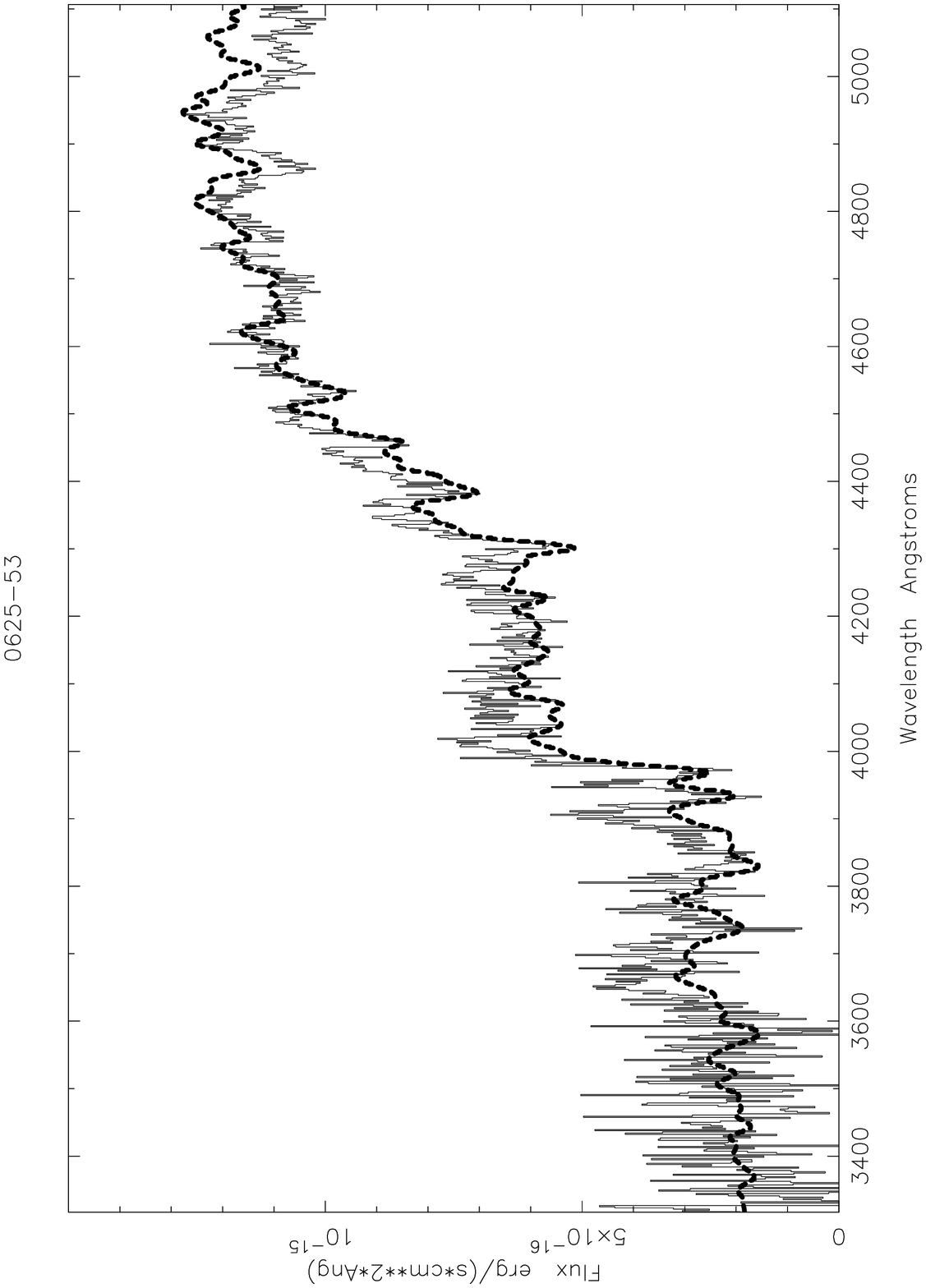}}
\put(0,0){\includegraphics{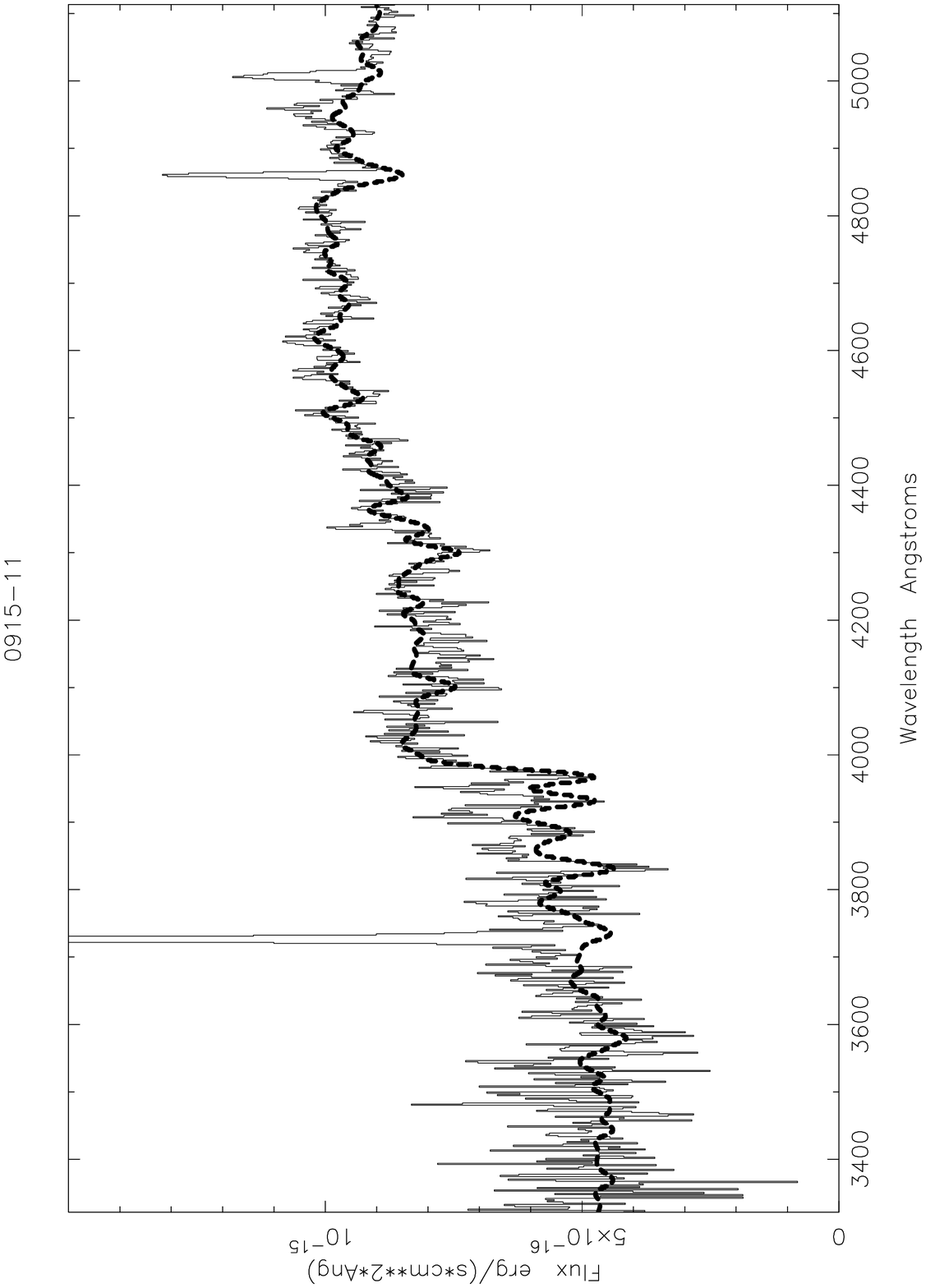}}
\end{picture}
\caption{The rest frame intensity spectra of the 13 observed radio
galaxies in the sample following nebular continuum subtraction (in the
cases of 1514+07 and Hydra A) and a correction for galactic reddening
(thin solid line) and the best-fitting one- or two-component model
(thick dotted line). See text and Table 3 for further details.}
\end{figure*}

\setcounter{figure}{0}
\begin{figure*}
\setlength{\unitlength}{1mm}
\label{fig1a.fig}
\begin{picture}(10,225)
\put(0,0){\includegraphics{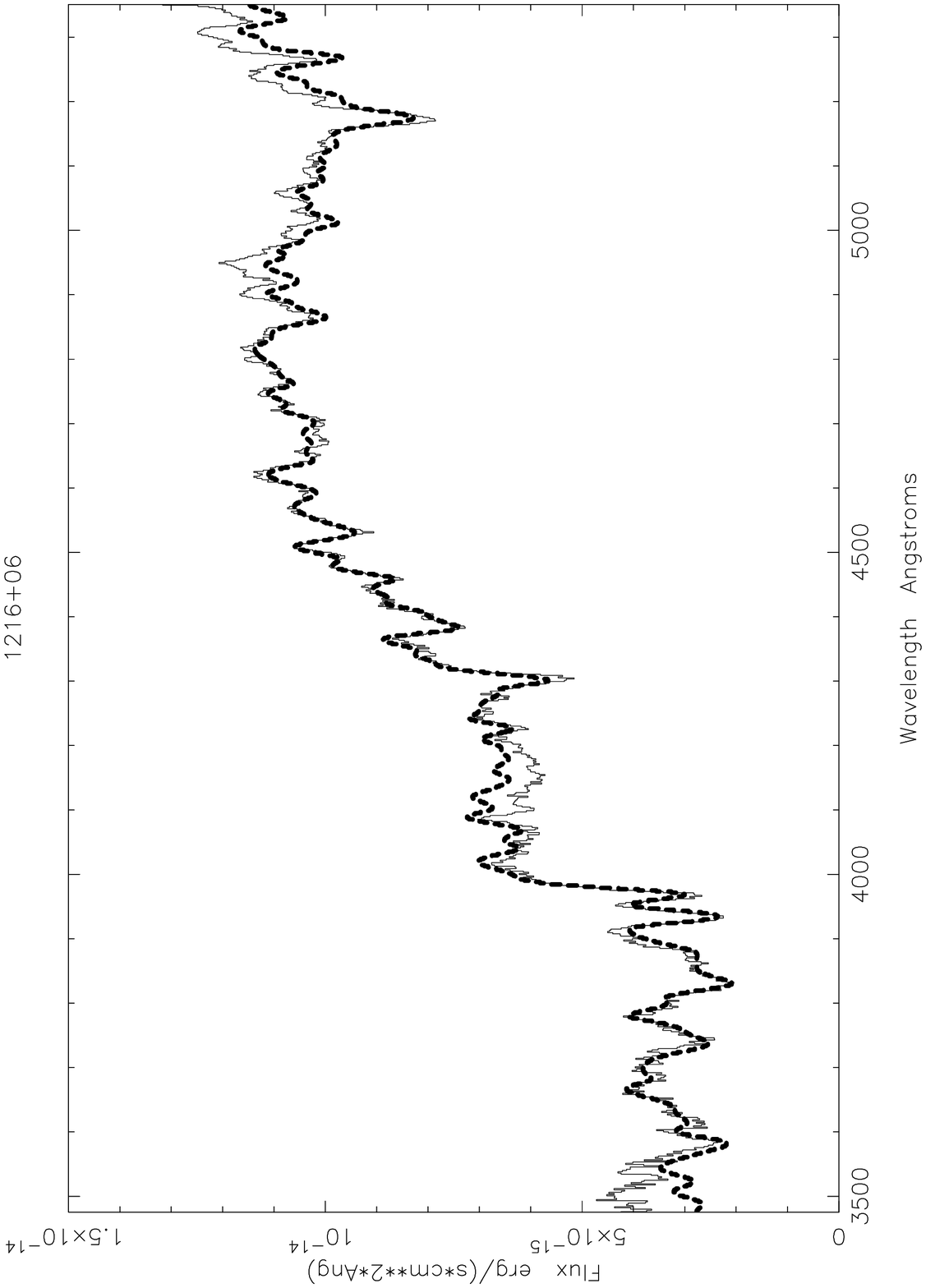}}
v\put(0,0){\includegraphics{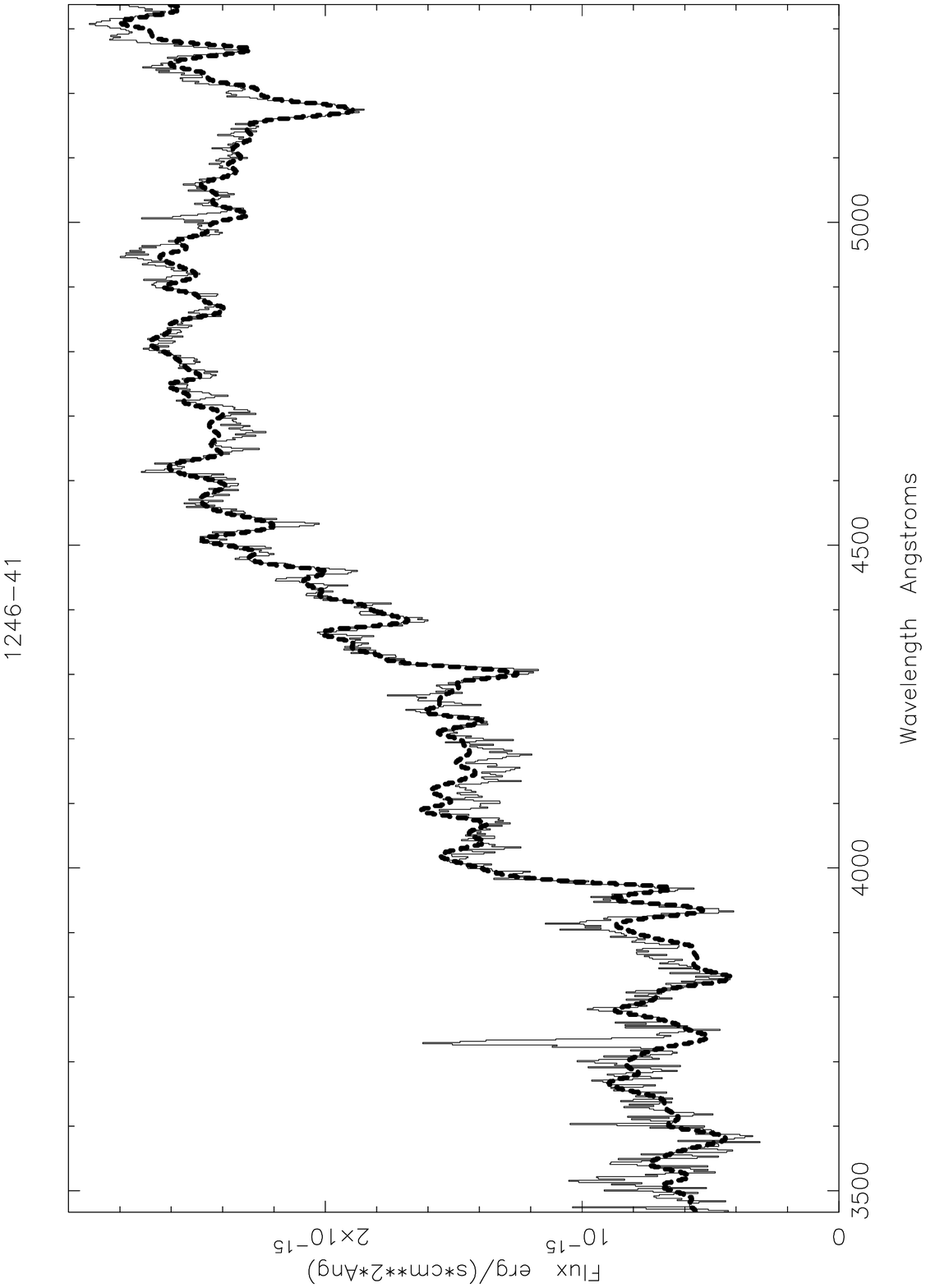}}
\put(0,0){\includegraphics{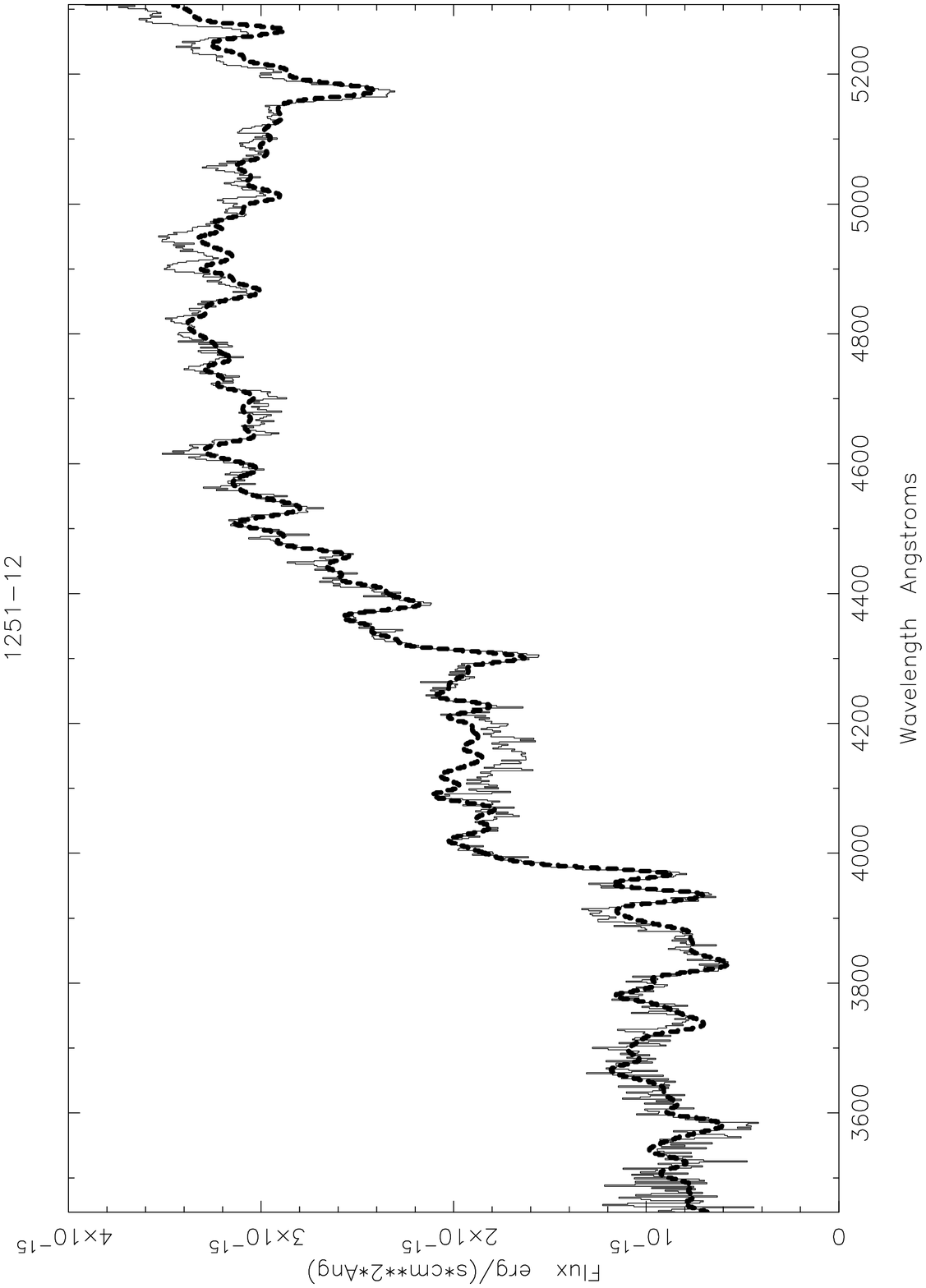}}
\put(0,0){\includegraphics{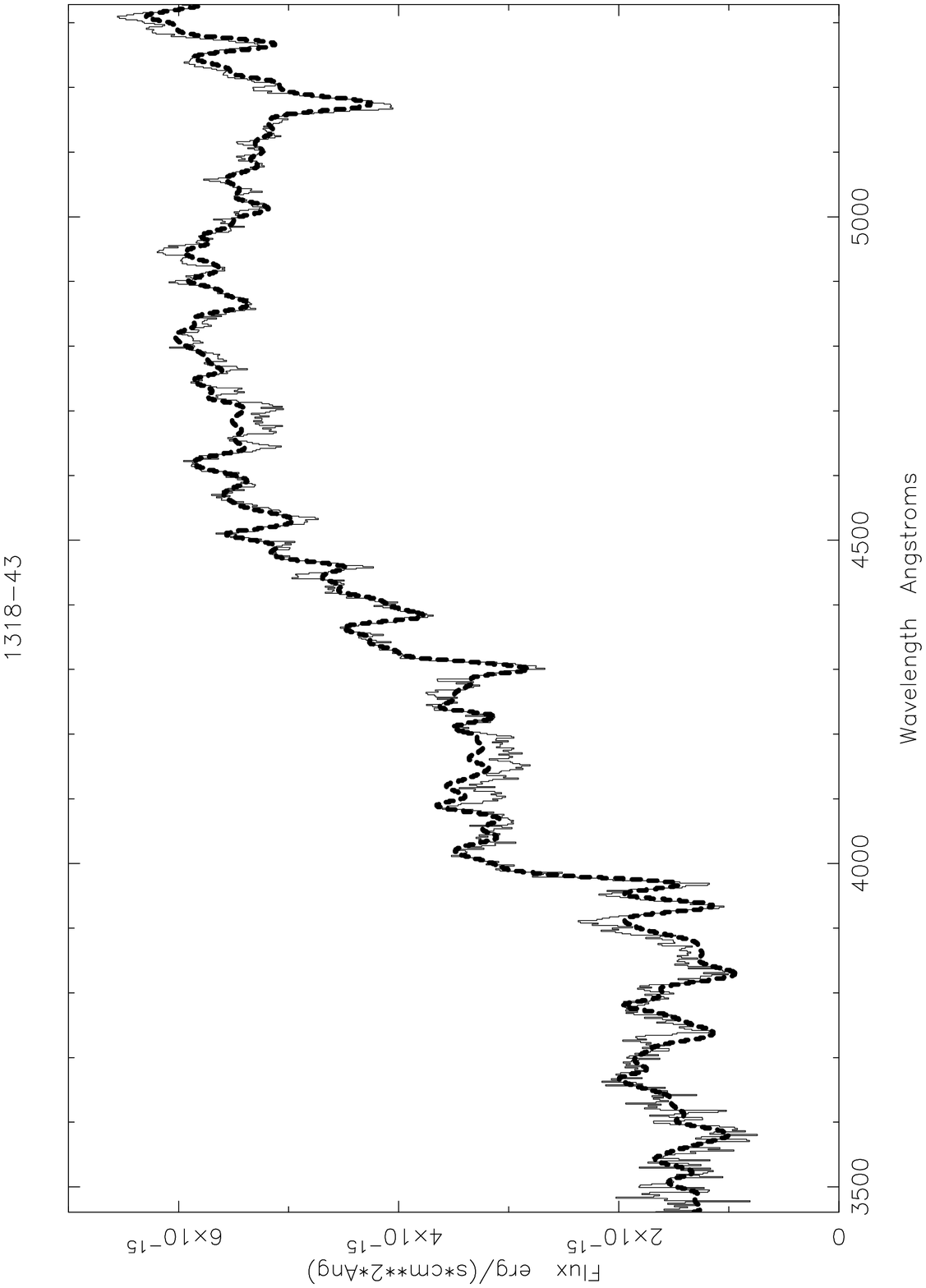}}
\put(0,0){\includegraphics{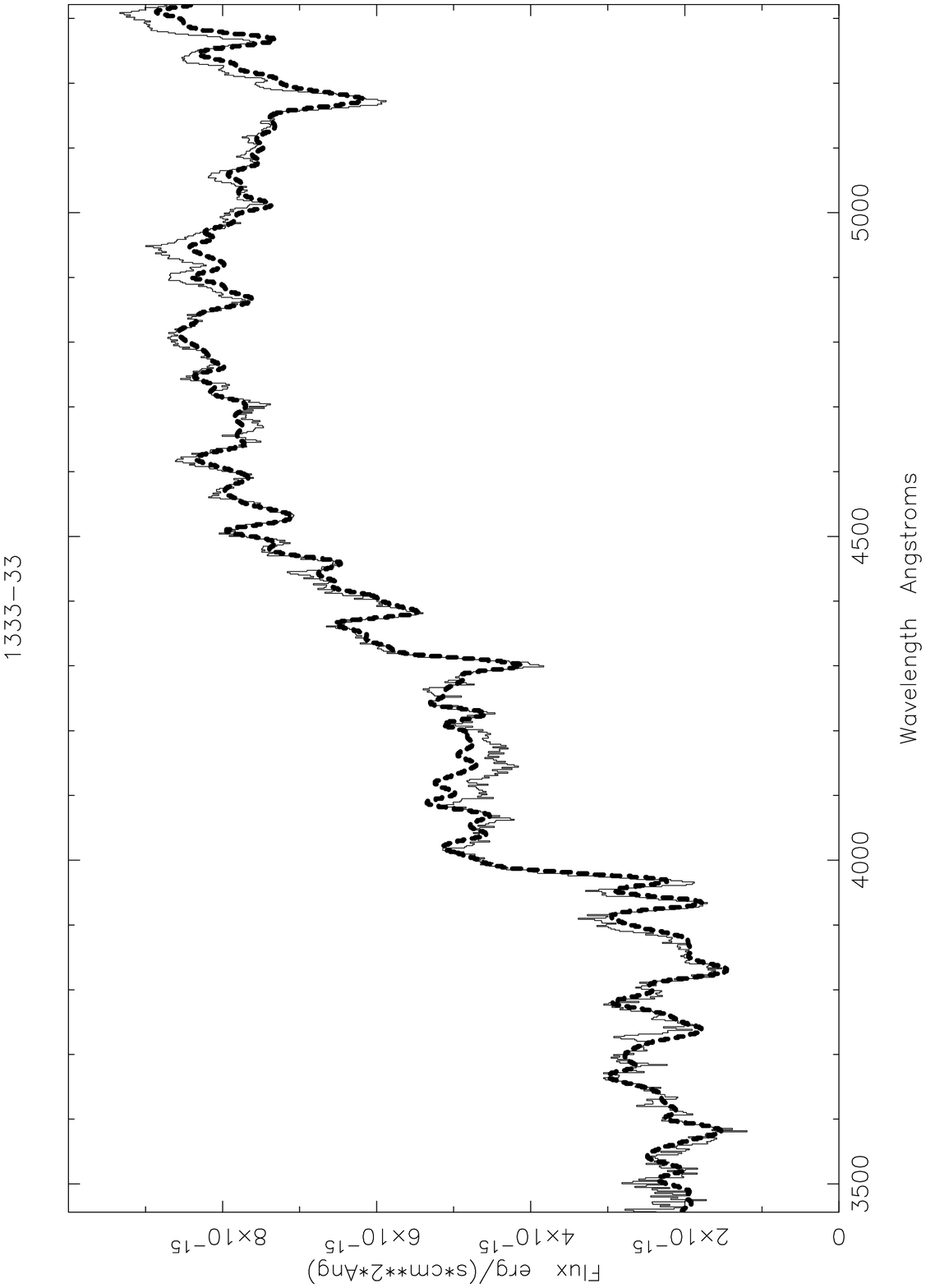}}
\put(0,0){\includegraphics{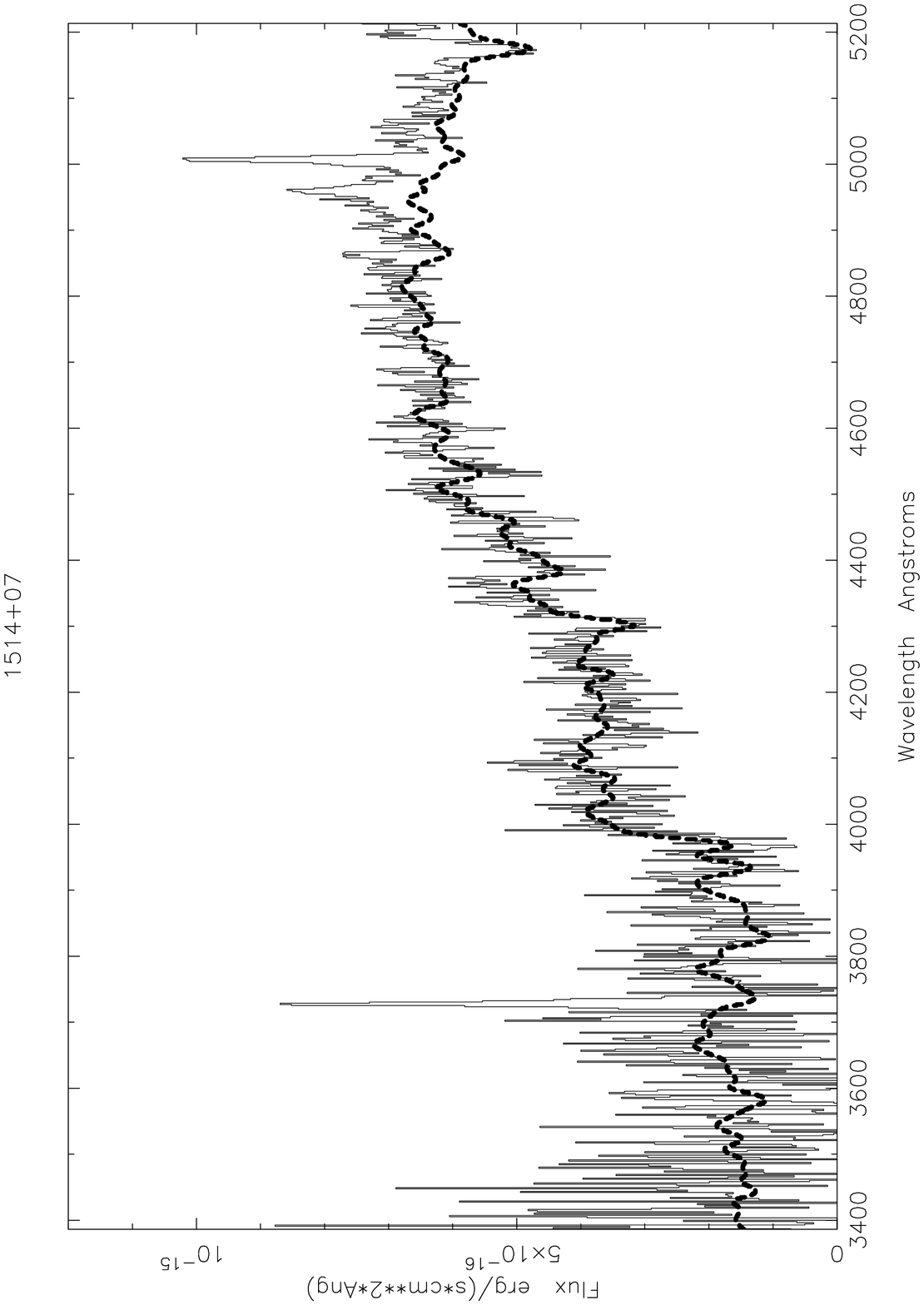}}
\end{picture}
\caption{cont.}
\end{figure*}

\setcounter{figure}{0}
\begin{figure*}
\setlength{\unitlength}{1mm}
\label{fig1b.fig}
\begin{picture}(10,70)
\put(0,0){\includegraphics{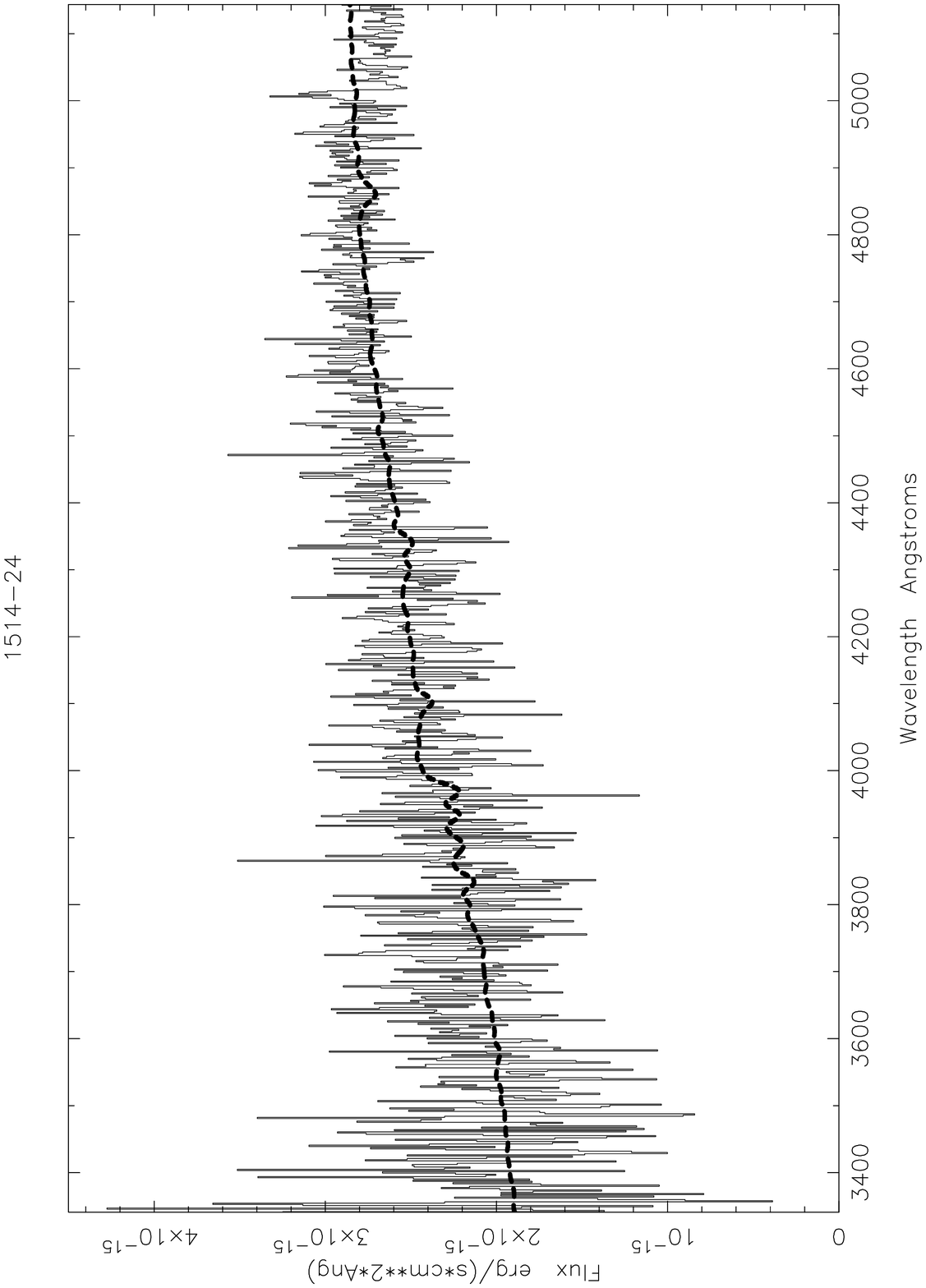}}
\end{picture}
\caption{cont.}
\end{figure*}

\section{Results}

We have used the extracted spectra (Figure 1) to derive both the
luminosity of the emission lines and the characteristics of the continuum.
The latter gives information on the presence of a possible UV excess and
on the stellar population.  As in the previous studies carried out on
powerful radio galaxies (Wills et al. 2002, Tadhunter et al. 2002), we
combine two different approaches to investigate the possible presence and
origin of a UV excess. We first measured the 4000 {\AA} break and then
performed a more detailed fit to the continuum. Furthermore, by
subtracting the best-fit to the continuum of the spectra, we were able to
detect emission lines otherwise too weak to be seen.

\subsection{4000 {\AA} break measurement}

A way to quantify the UV excess is to measure the 4000 {\AA} continuum
break (D(4000), Bruzual 1993). This is prominent in the spectra of
evolved stellar populations and therefore characterises the large
decrease, by a factor of $\sim$ 2 in the flux below 4000 {\AA}, in the
continuum emission of early-type galaxies. In line with our previous
work on powerful radio galaxies, we modify the bins used by Bruzual to
precisely define D$^\prime$(4000), in order to avoid emission line
contamination, and instead use the ratio of the total flux in a bin
100 {\AA} wide centred on 4200 {\AA} (rest frame) to the total flux in
a bin of identical width centred on 3800 {\AA}. Using this modified
definition, we can calculate that elliptical galaxies with solar
metallicity and ages of 10 and 15 Gyr have values of D$^\prime$(4000) of
approximately 2.1 and 2.3 respectively, and so any value significantly
less than these is indicative of an additional blue component.  In
Table~2 we show the D$^\prime$(4000) values for the sample.

Of the 13 observed radio galaxies, 8 have D$^\prime$(4000) $>$ 2.0 but
for 5 other galaxies the values are significantly less than this,
indicating the presence of an additional blue component. For example,
Ap~Lib shows a very low value of D$^\prime$(4000), consistent with it
being a BL~Lac. However, note that a low value of D$^\prime$(4000)
alone is not sufficient to classify a source as a BL~Lac and instead
the classification of BL~Lacs from our sample needs to be made on an
individual basis.

\subsection{Continuum modelling}

To learn more about the nature of the optical continua in the sources,
we have also modelled the continua using the results of published
isochrone spectral synthesis models.  The modelling of the continuum
was carried out using the method already described in previous papers
(e.g. see Tadhunter et al. 1996, Robinson et al. 2000, Wills et
al. 2002).  To model the continuum spectral energy distributions
(SEDs) of each galaxy we use the intensity spectra shown in Figure 1
(in the observed wavelength range 3400 to 5400 {\AA}, except for
0453-20 and 0620-52 where the observed wavelength range has been
cropped to 3700 - 5400 {\AA}), which have been corrected for galactic
reddening and where applicable, had the nebular continuum subtracted.
The continuum of each galaxy was initially modelled using two
components chosen to represent AGN light and the light of the host
elliptical galaxy. For the elliptical galaxy template spectrum we used
the Bruzual \& Charlot instantaneous burst models (see Leitherer et
al. 1996) with Salpeter (1955) IMF, solar metalicity and age ranges
from 15 to 10 Gyr. For the AGN light we use a power-law of the form
f$_{\lambda}$ $\propto$ $\lambda^{+\alpha}$. We also tried
incorporating a third component corresponding to a contribution from
young stars, with the starburst age ranging from 0.05 - 5 Gyr. The
models were generated by using a normalising continuum bin of 4806.3 -
4830.0 {\AA} and scaling the different model components such that the
total model flux in the normalising bin was less than 125\% of the
observed flux. As a result a series of models were created and the
best-fitting model was found using a reduced chi-squared test.

The results from this fitting are given in \S 3.4 and are summarised
in Table~3. In addition we also quote the results from Boisson et
al. (1989) for the fitting to the continuum of 0521-36. From the 14
sources in the sample we find that in 8 cases, we require only an old
stellar population (10-15 Gyr) to fit to the continuum. In 3 cases
(0453-20, 0620-52, 0915-11) an additional young population (0.05 - 2
Gyr) is also required in order to obtain a good fit to the
continuum. In the 3 remaining cases, a power-law is required in order
to obtain a good fit (the three BL~Lac type objects 0625-35, 0521-36
and 1514-24).

\subsection{Emission Lines}

The best fit to the continuum has been subtracted from the spectra in
order to be able to identify the emission lines. This method allows
detections of emission lines otherwise hidden by the strong thermal
continuum. With the exception of Hydra~A and 1514+07, we detected only
(at most) [OII] and [OIII].

In the cases where both [OIII]4959 {\AA} and [OIII]5007 {\AA} lines
are visible, the fit to these lines has been performed by: {\sl i)}
forcing the width of the [OIII]4959 {\AA} and [OIII]5007 {\AA} lines
to be equal; {\sl ii)} forcing their separation to be 48 {\AA}; {\sl
iii)} forcing the intensity of the [OIII]5007 {\AA} line to be 3 times
the 4959 {\AA} line. If only the [OIII]5007 {\AA} line is visible we
fitted to this as a single line. If a specific line is not visible, we
have calculated 3$\sigma$ upper limits to the flux and luminosity of
that line, obtained from an 11 {\AA} (the average linewidth) region.
Table 2 lists the line fluxes, luminosities and line ratios as
measured from the spectra. In addition we also quote the values for
0521-36 from Tadhunter et al. (1993) where available.

In 6 galaxies (0625-35, 1246-41, 1251-12 ([OII] and [OIII]); 1216+06,
1318-43, 1333-33 ([OII] only)) we detected emission lines when
previously (Tadhunter et al. 1993) we had just upper limits.  In
1514-24 we detect only [OIII], whereas in Tadhunter et al. both [OII]
and [OIII] were weakly detected. Note that two out of the three
objects which show a young stellar population component (the only
exception being Hydra~A) do not show strong emission lines (only upper
limits could be given)

The model subtracted spectra are shown in Figure 2.

\begin{figure*}
\setlength{\unitlength}{1mm}
\label{fig2.fig}
\begin{picture}(10,225)
\put(0,0){\includegraphics{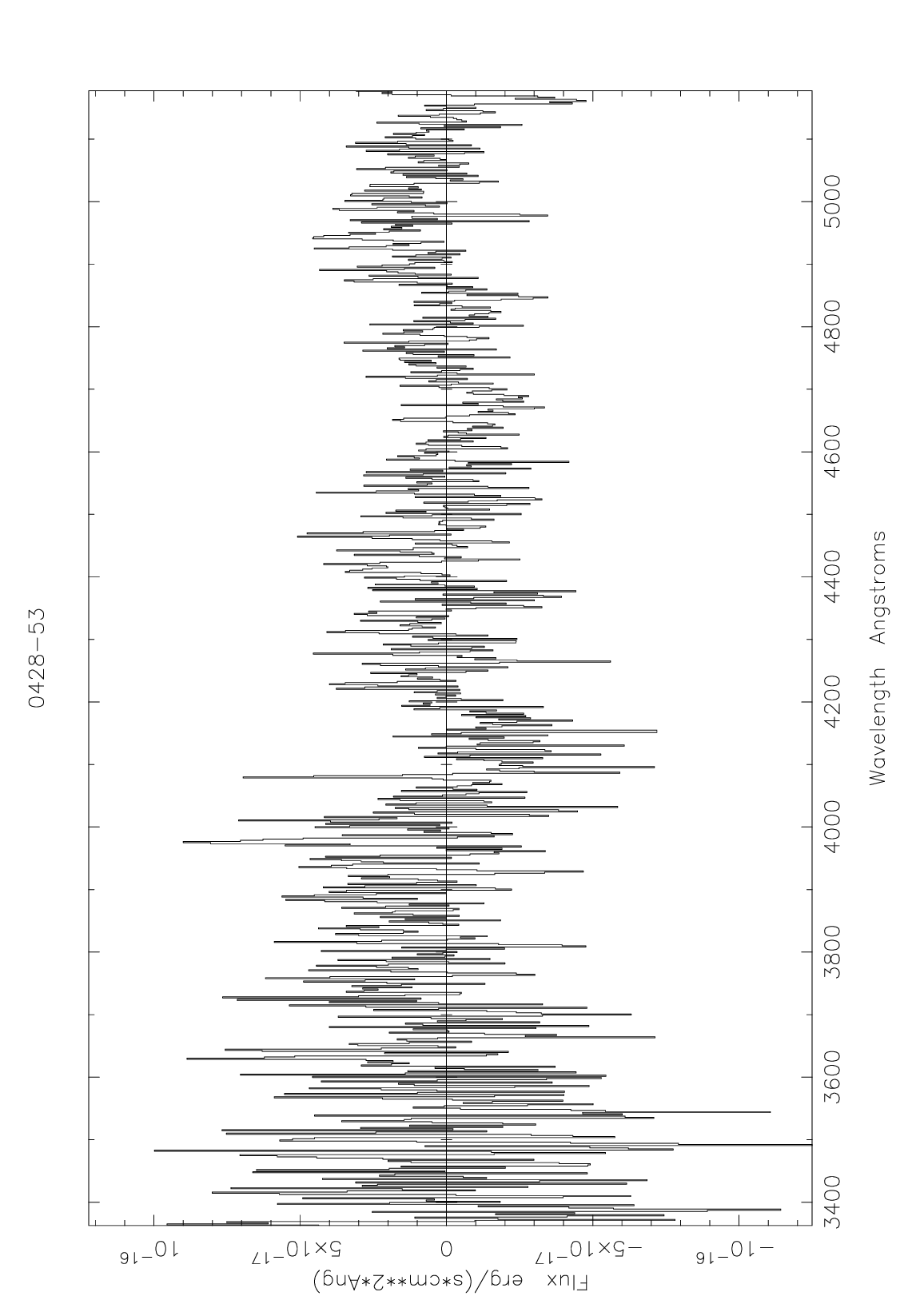}}
\put(0,0){\includegraphics{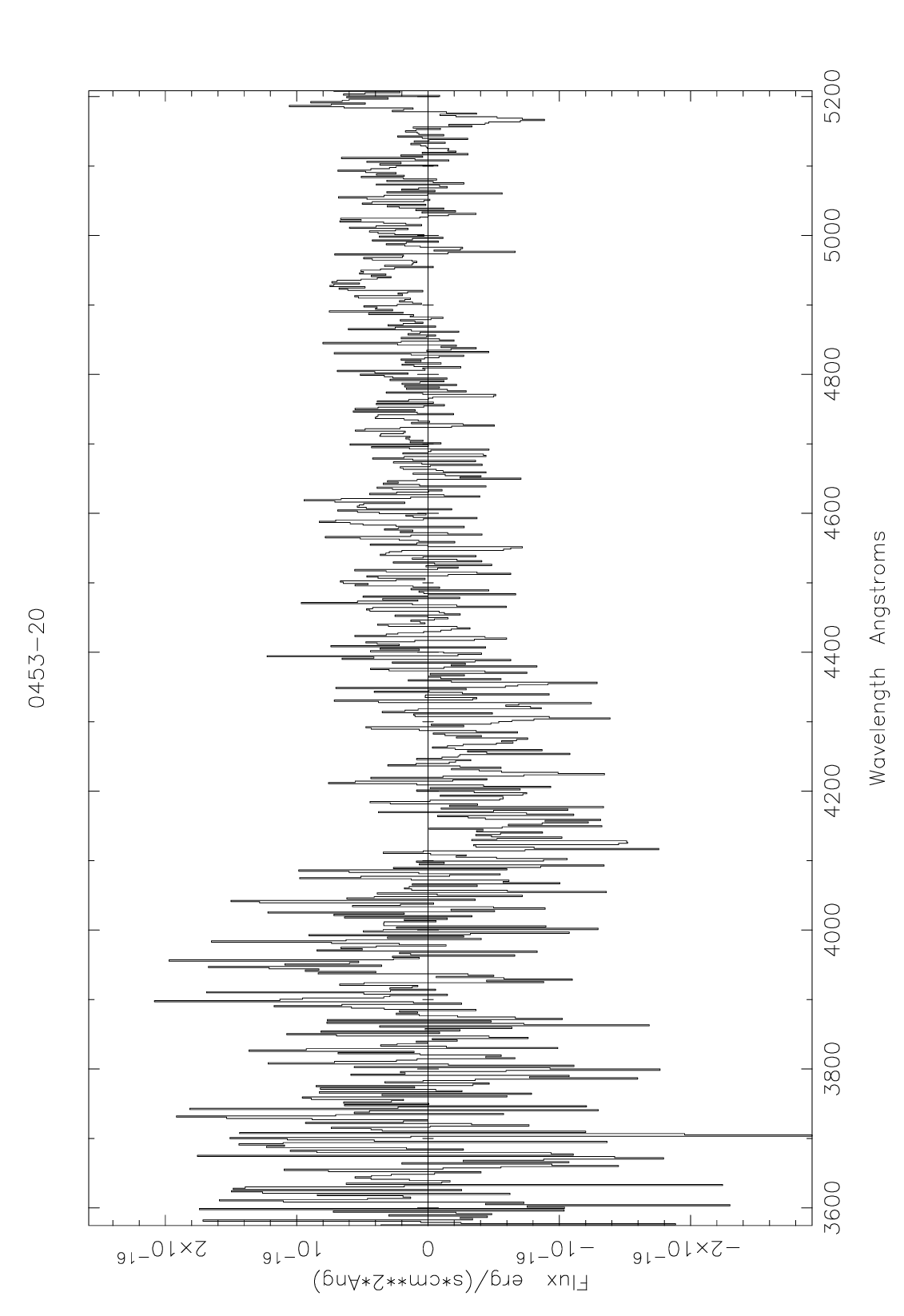}}
\put(0,0){\includegraphics{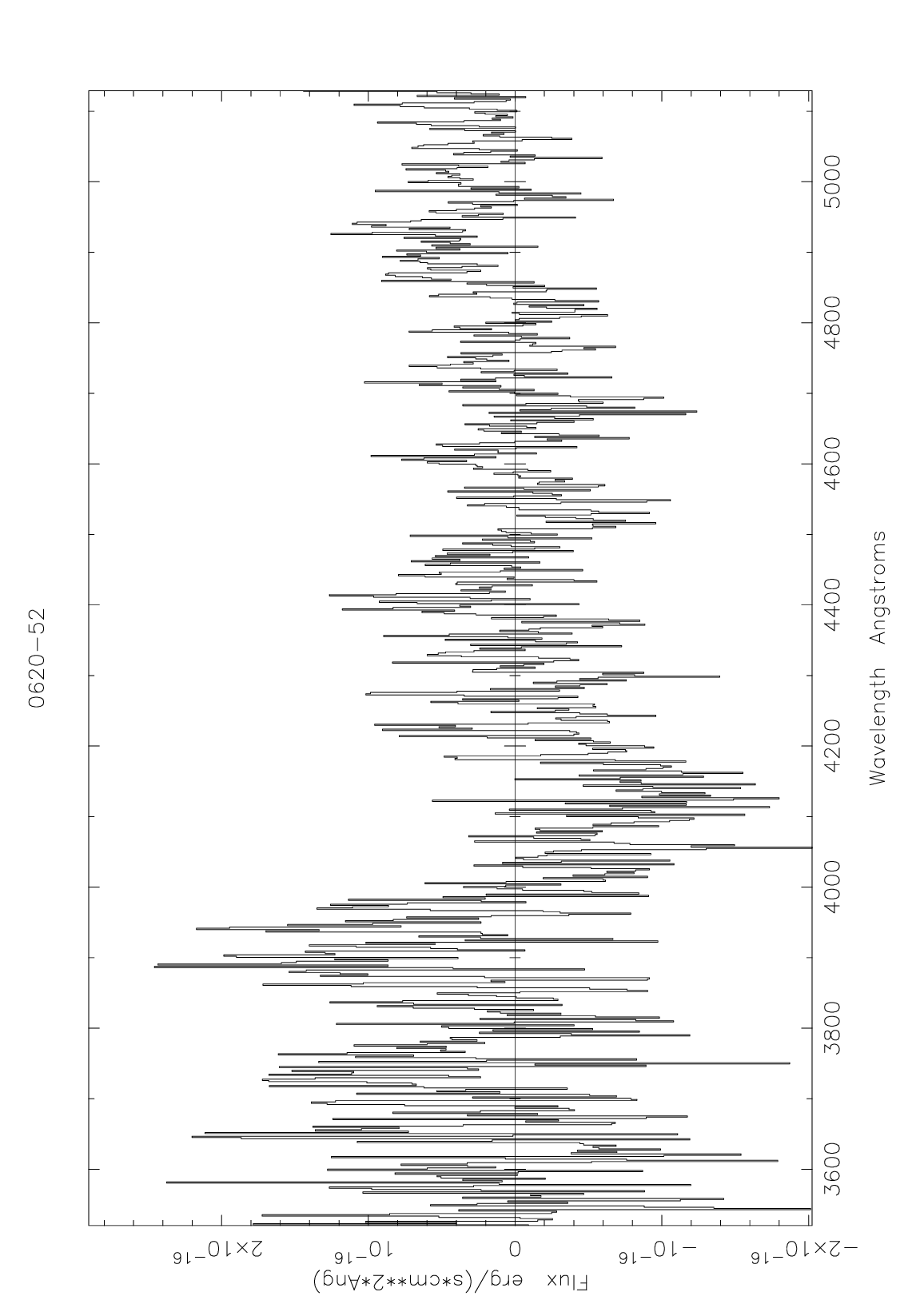}}
\put(0,0){\includegraphics{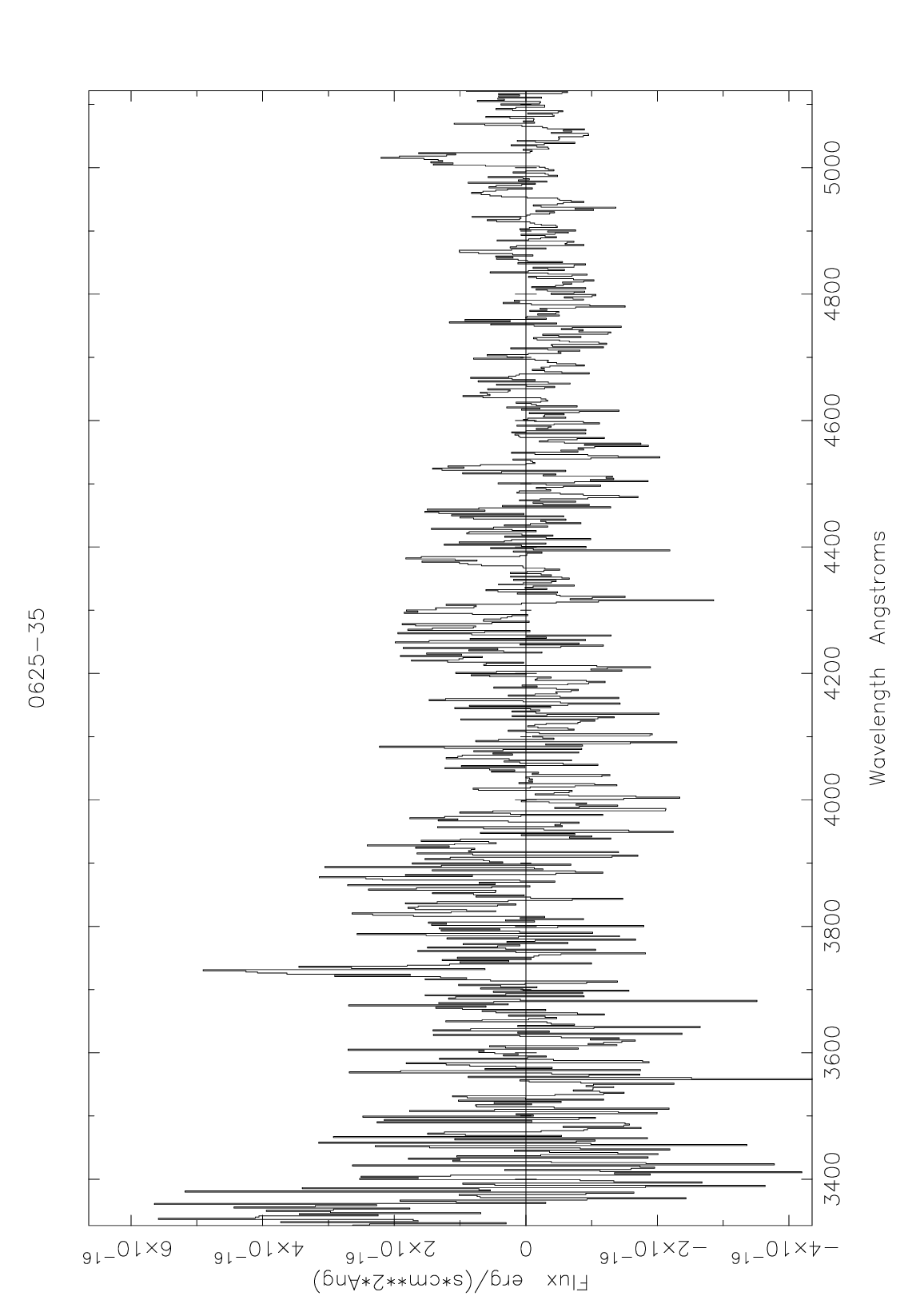}}
\put(0,0){\includegraphics{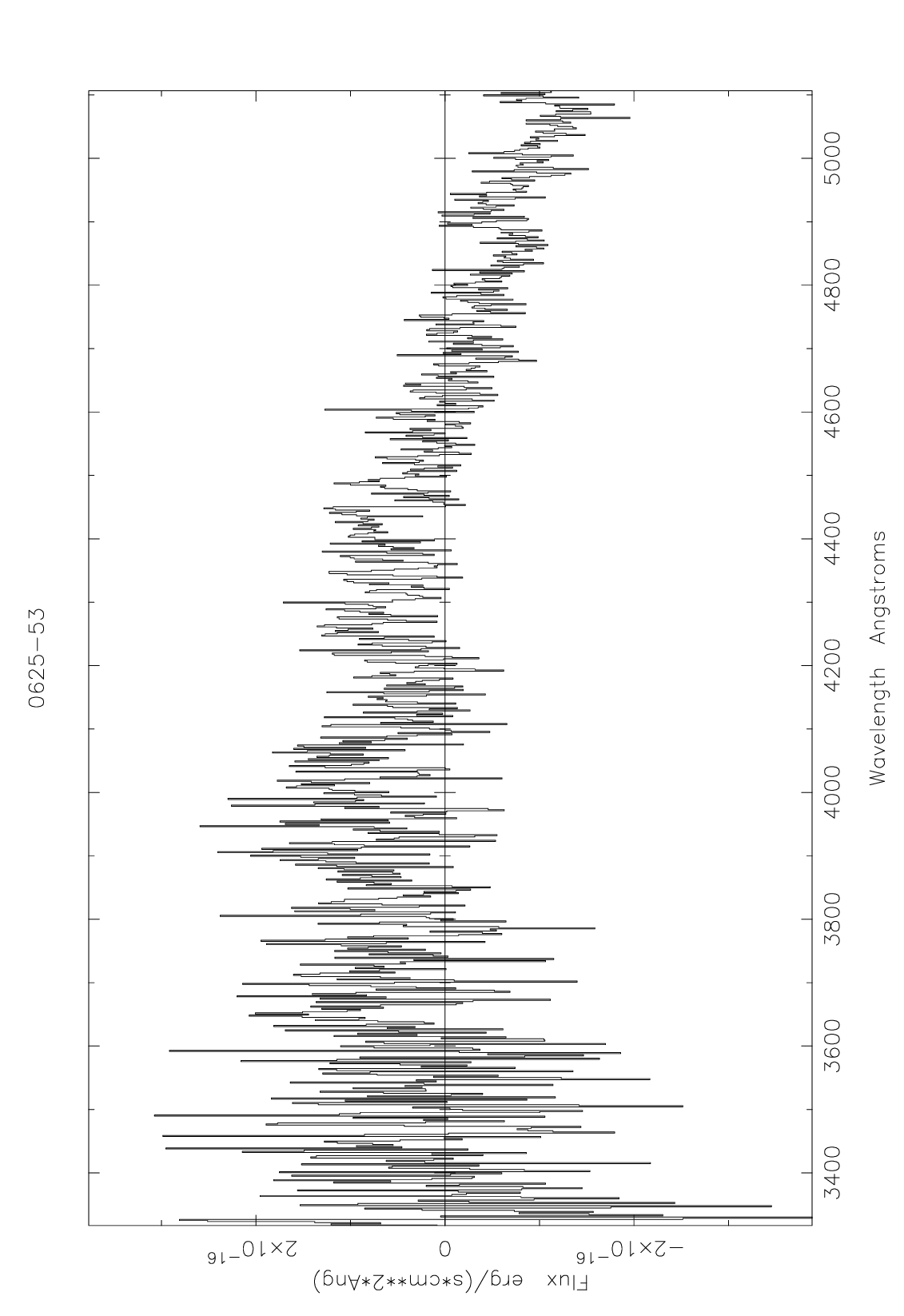}}
\put(0,0){\includegraphics{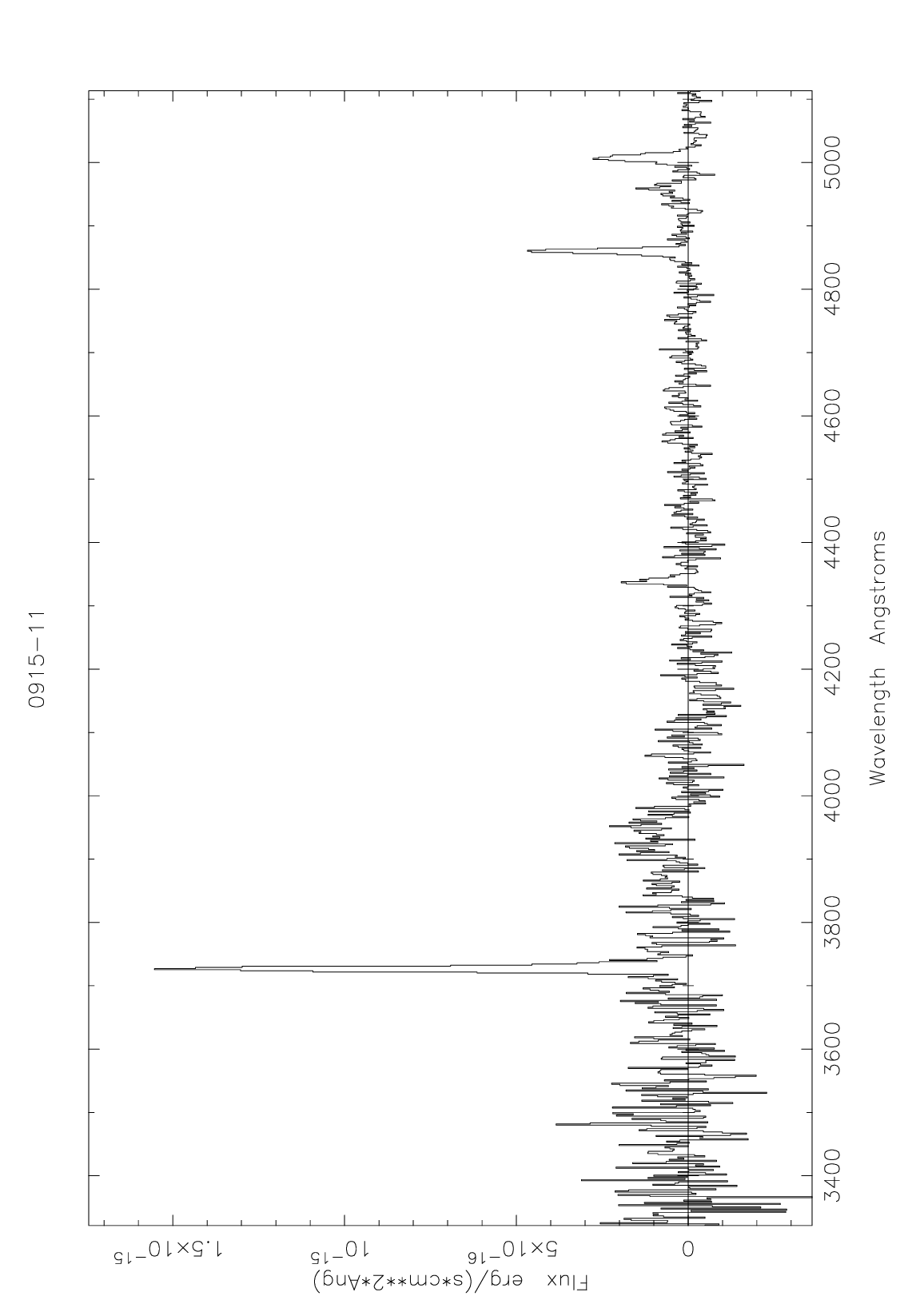}}
\end{picture}
\caption{The rest frame model-subtracted spectra of the 13 observed
radio galaxies in the sample. The best-fitting model for each galaxy,
as detailed in Table 3, has been subtracted.}
\end{figure*}

\setcounter{figure}{1}
\begin{figure*}
\setlength{\unitlength}{1mm}
\label{fig2a.fig}
\begin{picture}(10,225)
\put(0,0){\includegraphics{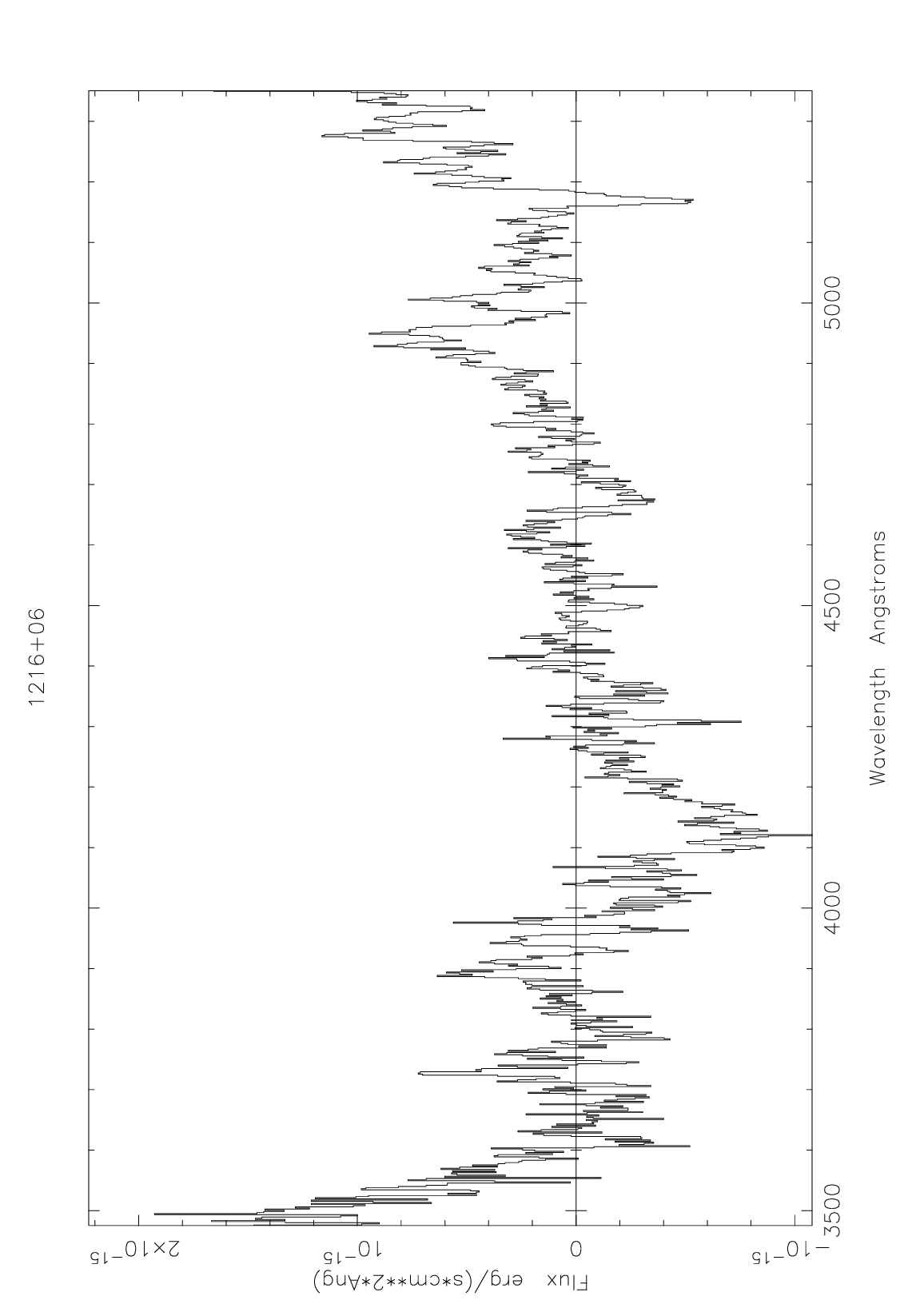}}
\put(0,0){\includegraphics{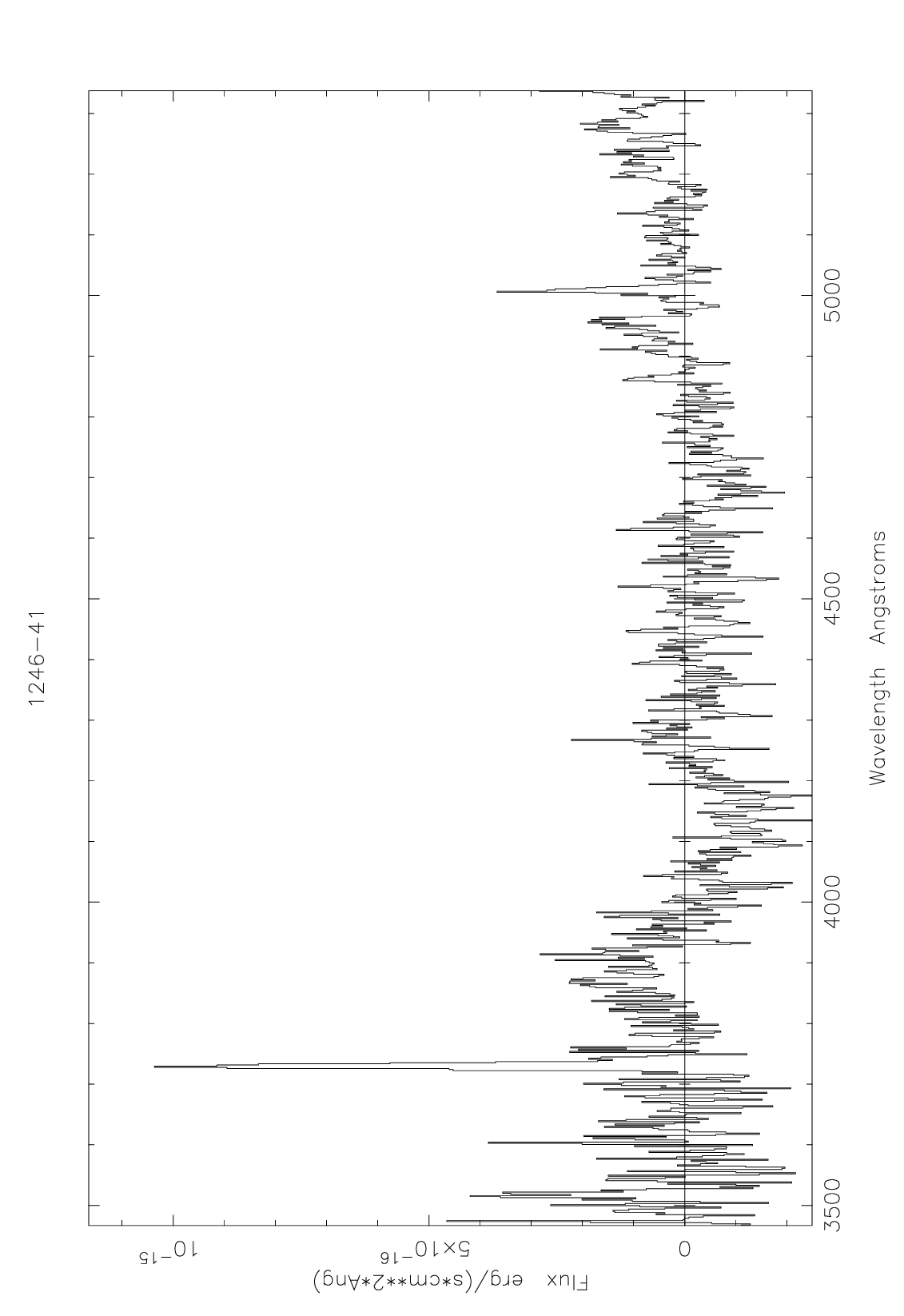}}
\put(0,0){\includegraphics{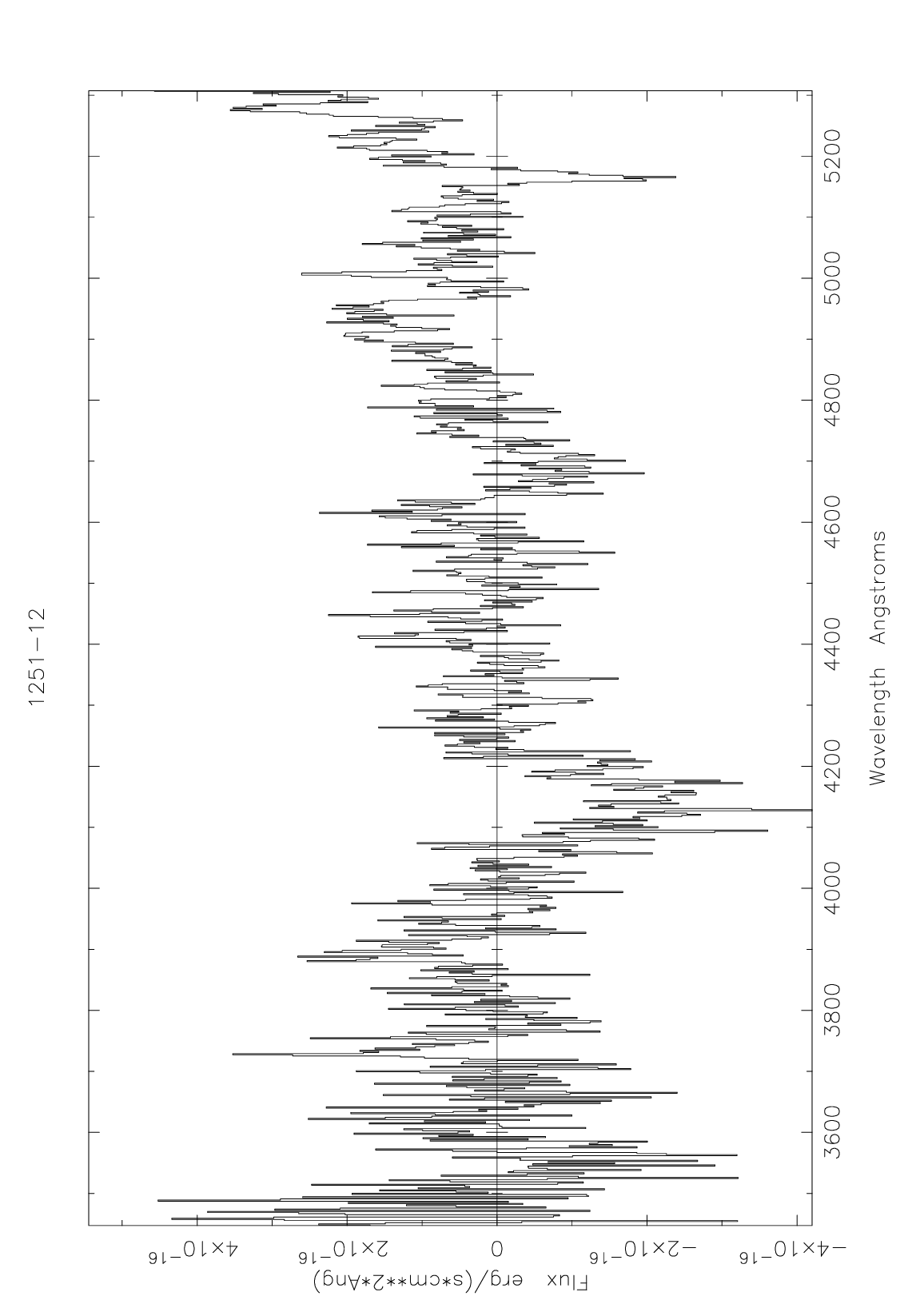}}
\put(0,0){\includegraphics{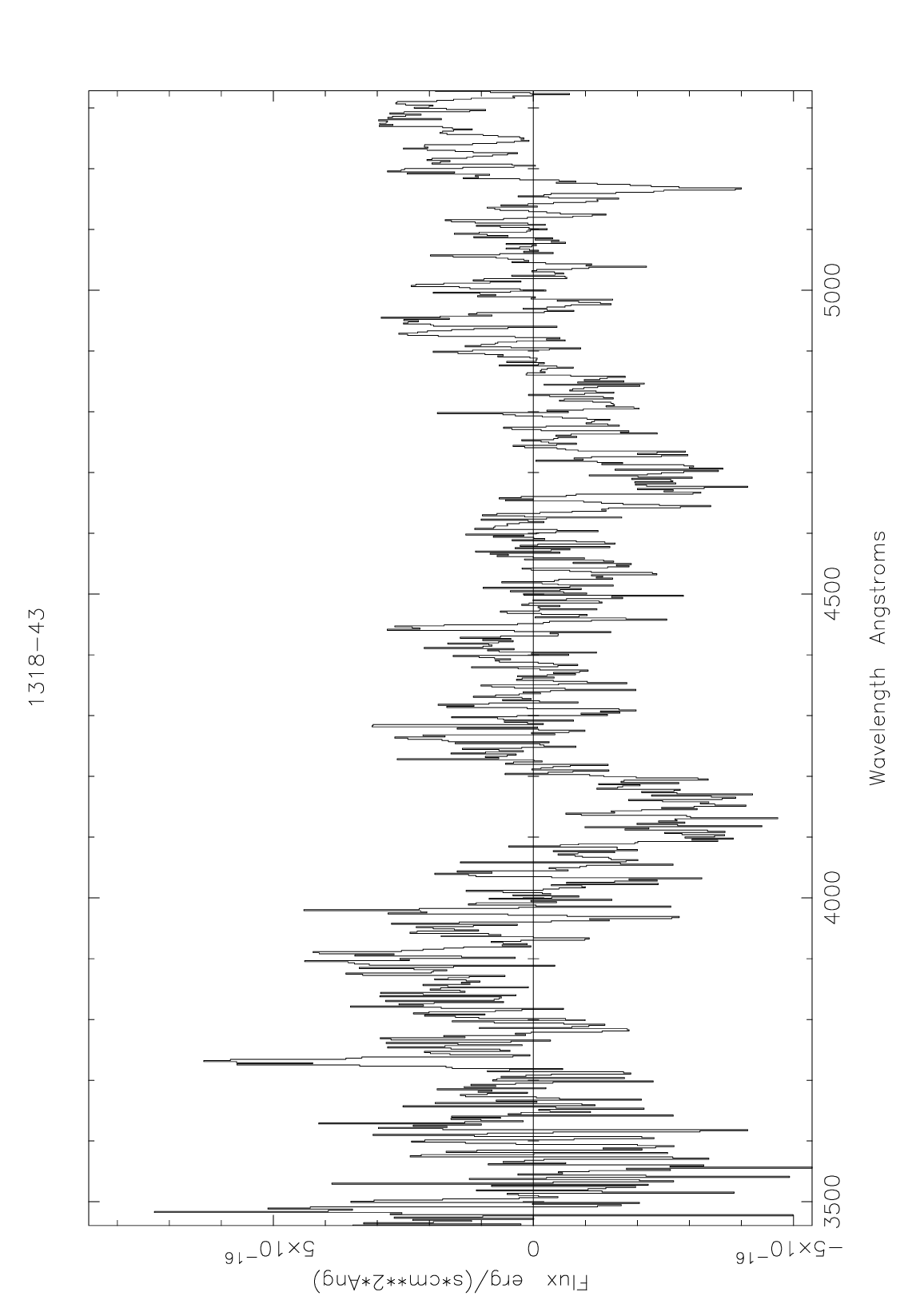}}
\put(0,0){\includegraphics{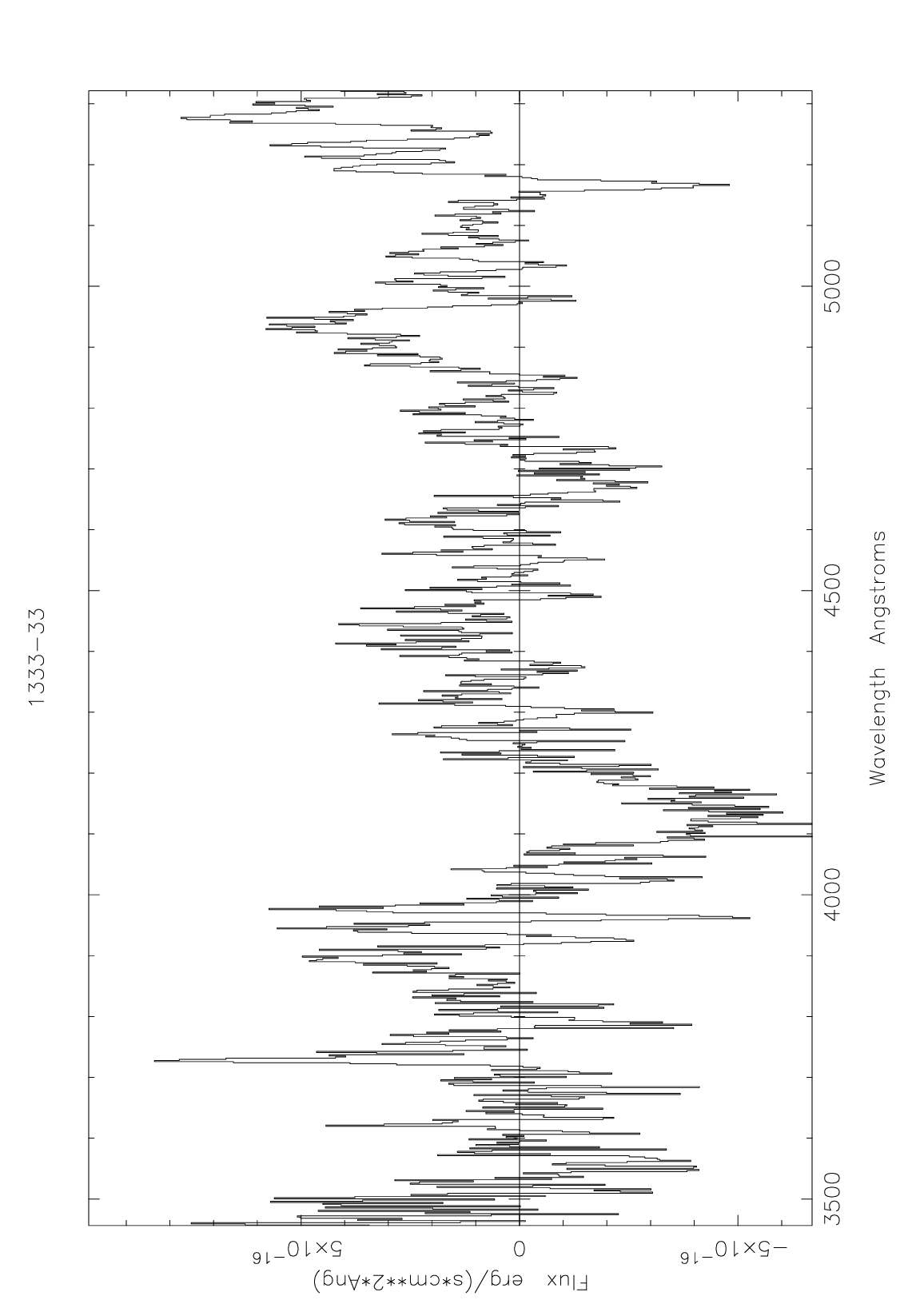}}
\put(0,0){\includegraphics{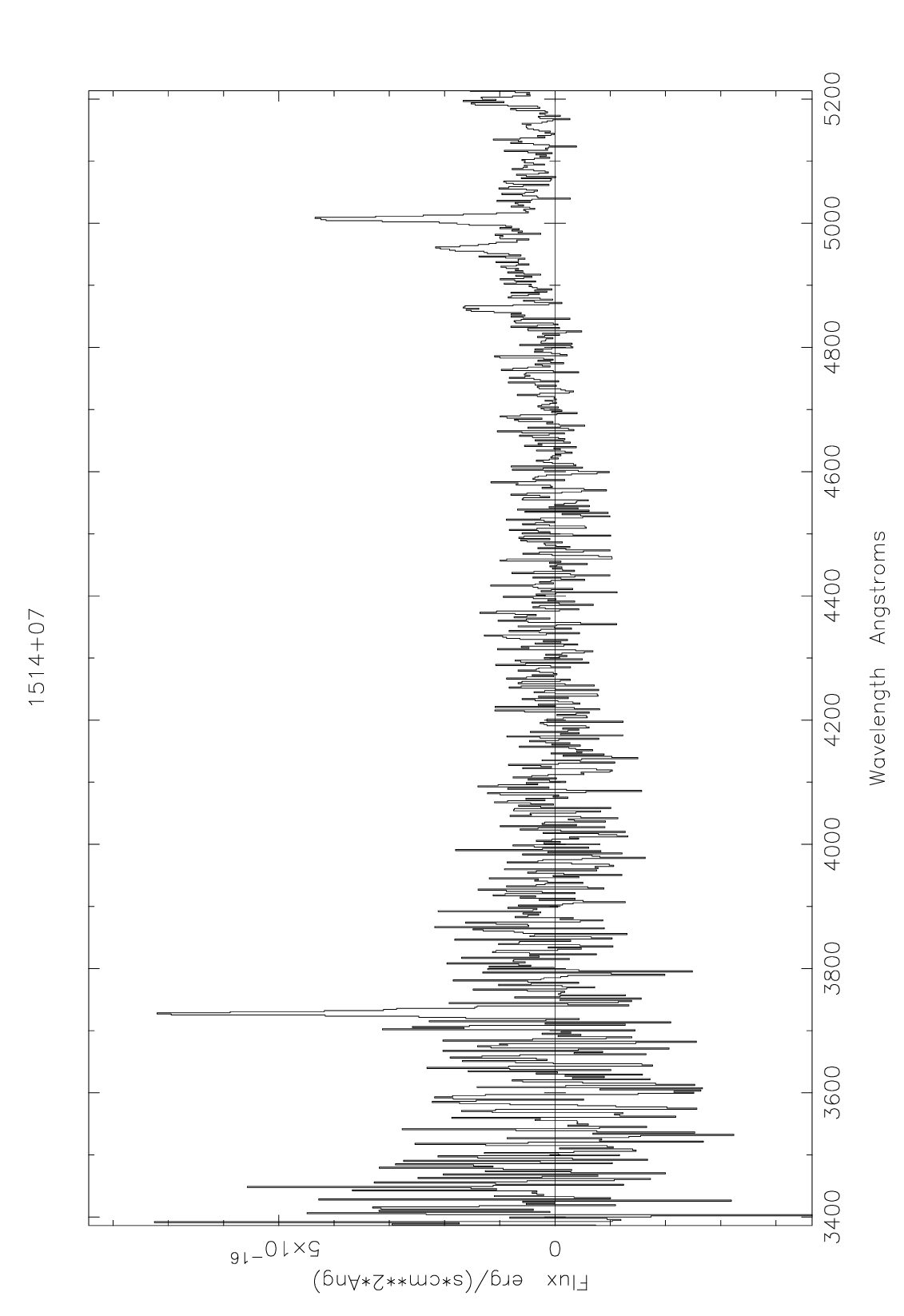}}
\end{picture}
\caption{cont.}
\end{figure*}

\setcounter{figure}{1}
\begin{figure*}
\setlength{\unitlength}{1mm}
\label{fig2b.fig}
\begin{picture}(10,70)
\put(0,0){\includegraphics{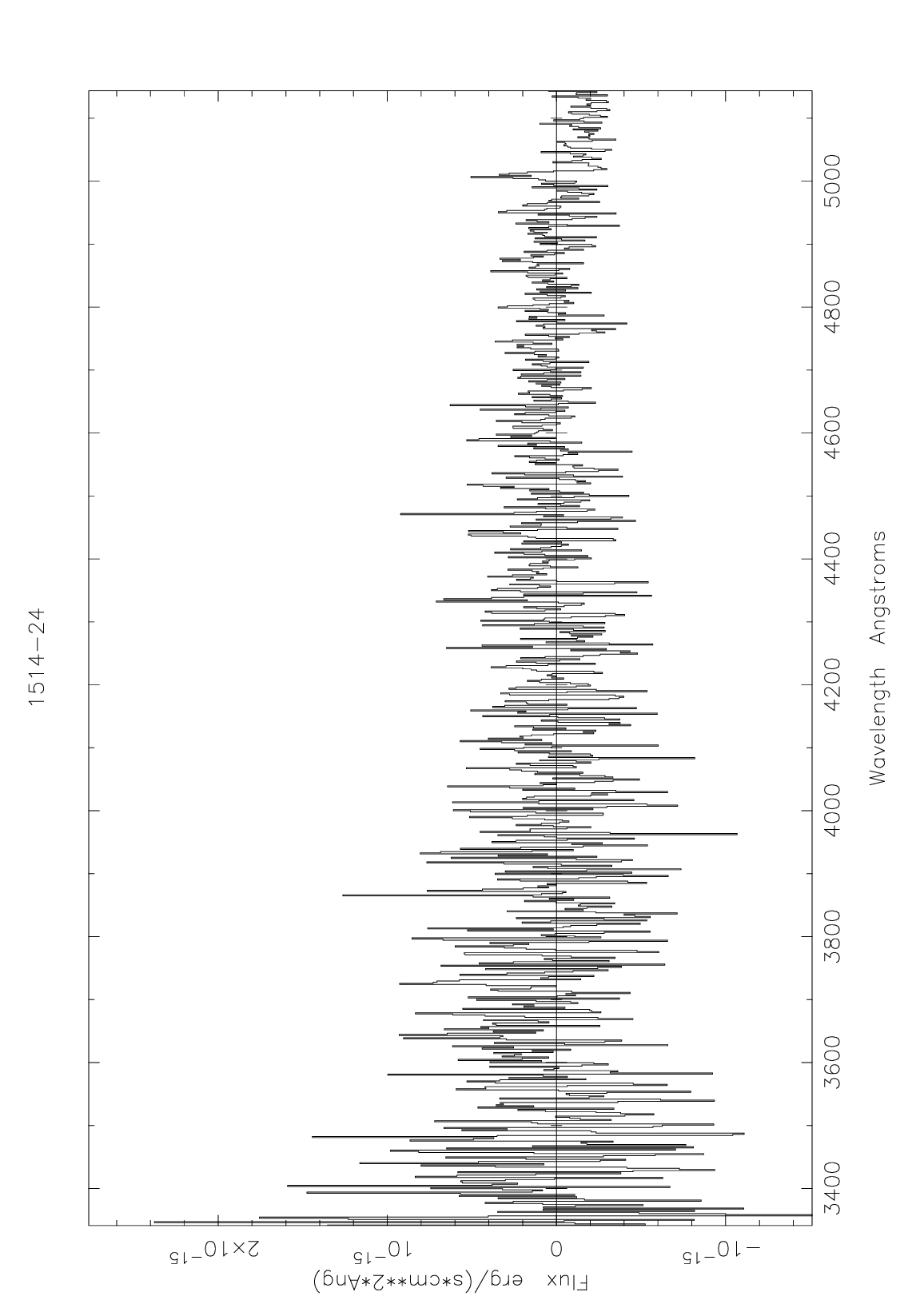}}
\end{picture}
\caption{cont.}
\end{figure*}

\begin{table*}
\centering
\caption{The best-fitting models to the continuum for each galaxy in
our sample. In the cases where more than one model is given, the first
one is the one used in the figures and in the deduction of the
emission line data. In many cases we actually find acceptable fits
with a range of young and old stellar ages, of which the one shown is
an example. More details are given in \S 3.4. The results for the
modelling of the continuum of 0521-36 are given from Boisson et
al. (1989).}
\label{tab3}
\begin{tabular}{clr}
\hline\hline\\
{\bf Name }   & {\bf Best fit}  & ${\bf \chi^2}$ \\
              &   \\
\hline    
0427-53 & 15 Gyr (101.5\% )                          & 1.44 \\
0453-20 & 15  Gyr  (88.8\%) + 0.05 Gyr (10.2\%)      & 0.597  \\
          & 5 Gyr (98.7\%)                      &  0.673  \\
0521-36 & Stellar spectrum of M31 + power-law ($\alpha$ = -0.5) & - \\
0620-52 & 15 Gyr (67.7\%) +  2 Gyr (33.4\%)         & 0.746   \\
0625-35 & 15 Gyr (45.8\%) + power-law ($\alpha$ = -1.15, 55.5\%) & 0.108 \\
0625-53 &  15 Gyr with E(B-V)=0.3 magnitudes of reddening(104\%) & 2.59  \\
0915-11 & 2 Gyr ( 89.3\%) +  0.05 Gyr ( 10.8\%)    & 0.835   \\
          & 10 Gyr (70.0\%) +  0.1 Gyr (30.0\%)      & 0.948   \\
1216+06 &  10 Gyr (99.4\%)                         & 0.617   \\
1246-41 &  15 Gyr (101.4\%)                        & 0.528   \\
1251-12 & 12.5 Gyr (99.0\%)                         & 0.889   \\
1318-43 &  15 Gyr (102\%)                          & 0.836  \\
1333-33 & 12.5 Gyr (99.9\%)                         & 0.939   \\
1514+07 &  15 Gyr (98.5\%)                         & 1.33    \\
1514-24 & 10 Gyr (17.6\%) +  power-law ($\alpha = 0.72$,  83.7\%) & 0.625  \\
\hline \\
\end{tabular}
\end{table*}

\subsection{Notes on individual objects} 

In 8 out of 14 of the objects in our sample we find no evidence for a UV
excess and instead, an old stellar population (10-15 Gyr) provides an
adequate fit to both the general shape and the detailed absorption line
features of the continuum spectra. For each of these objects, we can
achieve a marginally improved fit to the continuum if we include a small
contribution from a power-law component (typically $<$20\% of the flux in
the normalising bin). The luminosities of the required power-law
componenets (log L (W/Hz) = 19.8-20.5 in the normalising bin) are
consistent with those measured for the nuclear point source components
detected by Chiaberge et al. (2002, log L (W/Hz) = 18.1-20.3) in HST
images, albeit at the upper end of the range. However, the power-law
slopes are redder (typically $\alpha$ $>$ 4) than measured in most of the
Chiaberge et al. sources (typically $\alpha$ $\sim$ 2.5). Although we
cannot rule out the possibility that we are detecting the nuclear point
source components in our objects, an alternative possibility is that the
putative power-law component is compensating for inadequacies in the flux
calibration or the isochrone spectral synthesis models.

In 3 out of 14 of the objects in our sample we find evidence for a UV
excess and require an additional young stellar component (0.05 - 2
Gyr) to obtain an acceptable fit to the continuum. For two of these
objects (0453-20 and 0620-52) we find that the addition of a power-law
component to an old stellar population is also able to produce a good
fit to the continuum. However, the luminosities of the required
power-law components (log L (W/Hz) = 20.6-21.1 in the normalising bin)
are significantly larger than those of the point source components of
Chiaberge et al. (2002). In any case, although the addition of the
power-law component provides a good overall fit to the general shape
of the continuum spectra, the fits to the continuum SED and in
particular the detailed absorption line features, are significantly
improved when the power-law component is replaced by a young stellar
component. In the third object of this group (0915-11) the addition of
a power-law component does not provide an adequate fit to the
continuum, as is discussed in detail below.

In general, the models provide an excellent fit over most of the
spectral range covered by the data. However, we note that in the rest
frame wavelength range 4050 -- 4200 {\AA}, the models often show a
significant excess relative to the data (see Figures 1 and 2). Since
this excess is seen in the same rest wavelength range for objects
covering a wide range in redshift, we suggest that the excess
represents a problem with the models rather than with the flux
calibration.

\noindent
{\bf 0427-53} This galaxy does not show a UV excess either from the
4000 {\AA} break or from the fit to the continuum.  The best fit was
obtained by using a value of the redshift of z=0.0412 (z=0.038
previously quoted by Tadhunter et al. 1993) and with an old stellar
population only (15 Gyr gives $\chi^2$ = 1.44). No emission lines were
detected.

\noindent
{\bf 0453-20} This galaxy shows a UV excess (from D$^\prime$).  In
the fit to the continuum, a definite improvement is seen by the
addition of a young stellar component (e.g. $\chi^2$ is $\sim$ 3 for a
15 Gyr component only compared with $\sim$ 0.6 for a 15 Gyr component
plus a 0.05 Gyr component). The best fit is obtained with a 15 Gyr
component plus a 0.05 Gyr component which fits very well except for a
slight excess around 4100 {\AA} ($\chi^2$ = 0.597). In addition, we
also get good fits ($\chi^2$ $\sim$ 0.7) with a 15 Gyr component plus
a 0.1 Gyr component, or a 15 Gyr component plus a 2-5 Gyr component,
or a 5 Gyr component only. No emission lines were detected.

\noindent {\bf 0521-36} This source has an optically variable continuum
which has resulted in its classification as a BL~Lac object (see Danziger
et al. 1979). Both Danziger at al. (1979) and Boisson et al.
(1989) note that the optical SED of this source requires a strong
power-law component with $\alpha \sim$ -0.5, in addition to an elliptical
galaxy component. Both the [OII] and [OIII] lines were detected by
Tadhunter et al. (1993).

\noindent {\bf 0620-52} This galaxy shows a UV excess.  In the fit to the
continuum, a definite improvement is seen by the addition of a young
stellar component (e.g. $\chi^2$ is $\sim$ 2 for a 15 Gyr component only
compared with $\sim$ 0.7 for a 15 Gyr component plus a 2 Gyr component).
The best fit is obtained with a 15 Gyr component plus a 2 Gyr component
($\chi^2$ = 0.746), but adequate fits ($\chi^2$ $\sim$ 0.8) can also be
obtained for young stellar components of ages 1-5 Gyr and 0.05-0.1 Gyr.
This object was observed with the SAX satellite but has no obvious
non-thermal component (Trussoni et al. 1999).  No emission lines were
detected.

\noindent {\bf 0625-35} We suggest that this source could be a BL Lac
object on the basis of the results of the 4000 {\AA} break
(D$^\prime$(4000)=1.06) and from the fit to the continuum. It is not
possible to gain an adequate fit to the continuum with either an old
elliptical galaxy component or a young stellar component or a combination
of the two. Instead, a power-law component is required to gain a good fit
to the continuum, with the best fit being the combination of this
power-law with a 15 Gyr component ($\chi^2$ = 0.108). However, adequate
fits ($\chi^2$ $\sim$ 0.1) are also obtained with the combination of
power-law component and elliptical galaxy ages of 5-15 Gyr. Both the [OII]
and [OIII] lines are detected and the redshift derived from the fit of the
[OII] line (z=0.525) has been used in the fitting of the continuum
(z=0.055 previously quoted by Tadhunter et al. 1993).  Interestingly, in
this galaxy non-thermal emission has been detected in hard X-rays from SAX
(Trussoni et al. 1999). These results are consistent with the relatively
large R parameter reported for this source (see Table 1): 0625-35 has the
second largest R parameter in the group, after the well-known BL~Lac
object 1514-24. As further evidence for the orientation of this object,
the R-band imaging paper of Govoni et al. (2000) suggests the presence of
a nuclear point source.

\noindent {\bf 0625-53} No obvious UV excess is observed in this galaxy.  
The best-fitting model, although far from ideal ($\chi^2$ = 6.73), was
obtained using a 15 Gyr elliptical galaxy component only. The fit could
not improved by the addition of a young stellar component. However, the
fit was improved by reddening the elliptical galaxy component, in order to
represent intrinsic reddening within the source. The best fit was obtained
with E(B-V)=0.3 and although this final fit is still not ideal ($\chi^2$ =
2.59), the reddening offers a significant improvement. We used a value of
the redshift of z=0.0556 (z=0.054 previously quoted by Tadhunter et al.
1993). No emission lines were detected.

\noindent {\bf 0915-11 (Hydra~A)} This galaxy has a clear UV excess. It is
not possible to gain an adequate fit to the continuum with either an old
elliptical galaxy component ($\chi^2$ $>$ 14) or a young stellar component
alone ($\chi^2$ $>$ 2). The fit is somewhat improved with the addition of
a power-law component although is still far from ideal (e.g. 15 Gyr +
power-law with $\alpha$ = 0.88 has $\chi^2$ = 1.85). A young stellar
component is clearly required and in fact the best-fitting model comprises
a 2 Gyr component with a 0.05 Gyr component ($\chi^2$ = 0.835). Acceptable
fits ($\chi^2$ $\sim$ 0.9) are also obtained with a 2-15 Gyr component
combined with a 0.05-0.1 Gyr component. The presence of a young stellar
population in this galaxy has already been discussed by Melnick et al.
(1997) and Aretxaga et al. (2001). Melnick et al. state that the prominent
Balmer absorption and blue colour of the central disk suggest that massive
star formation activity has occurred over at least the past 0.01 Gyr.
Aretxaga et al. use their measured Balmer break index and stellar
evolutionary tracks to suggest that the last burst of star formation
occurred in the last 0.007-0.04 Gyr. In view of these particularly young
ages suggested for the most recent burst of star formation, we also
attempted to fit to the continuum of HydraA using a 0.01 Gyr component. We
find that, we can obtain an acceptable fit if the 0.01 Gyr component is
combined with a 2 Gyr component ($\chi^2$ = 0.924). The 2 Gyr + 0.05 Gyr
model still remains the best-fit to the continuum although there is little
difference between any of the acceptable fits. In conclusion, our results
are consistent with the young stellar component identified by Aretxaga et
al. We used a value of the redshift of z=0.0542 (z=0.054 previously quoted
by Tadhunter et al. 1993).

\noindent {\bf 1216+06 (3C270)} In this galaxy there is no obvious UV
excess. The best fit to the continuum is obtained using a redshift of
z=0.00747 (z=0.006 previously quoted by Tadhunter et al. 1993) and an old
stellar population (10 Gyr is best , 12.5 Gyr is also acceptable, both
with $\chi^2$ $\sim$ 0.6) only. There is no improvement to the fit by the
addition of a young stellar component. The [OII] line was marginally
detected.

\noindent {\bf 1246-41 (NGC~4696)} No obvious UV excess is detected for
this galaxy. The best fit to the continuum is with a 15 Gyr elliptical
component alone ($\chi^2$ = 0.528).  We find no improvement to the fit by
the addition of a young stellar component. We detect both the [OII] and
[OIII] lines and the redshift derived from the absorption and emission
lines is z = 0.00994 (z=0.009 previously quoted by Tadhunter et al. 1993).

\noindent {\bf 1251-12 (3C~278)} No obvious UV excess is detected in this
galaxy.  A reasonable fit to the continuum is obtained using an old
stellar component only (12.5 Gyr is best , 15 Gyr is also acceptable, both
with $\chi^2$ $\sim$ 0.8). There is no significant improvement to the fit
by the addition of a young stellar component.  We detect both the [OII]
and [OIII] lines and the redshift derived from the absorption and emission
lines is z = 0.0157 (z=0.015 previously quoted by Tadhunter et al. 1993).

\noindent
{\bf 1318-43 (NGC~5090)} There is no obvious UV excess in this
galaxy.  The best fit to the continuum is obtained with just an old
stellar population (15 Gyr, $\chi^2$ = 0.836). There is no significant
improvement to the fit with the addition of a young stellar
component. The [OII] line is detected and the redshift derived from
the absorption lines in the galaxy is z = 0.0117 (z=0.011 previously 
quoted by Tadhunter et al. 1993).

\noindent
{\bf 1333-33 (IC~4296)} There is no obvious UV excess detected in
this source.  The best fit to the continuum is achieved with just an
old stellar population (12.5 Gyr, $\chi^2$ = 0.939). The fit is not
improved with the addition of a young stellar component. Only the
[OII] line is detected.

\noindent {\bf 1514+07 (3C~317)} There is no obvious UV excess in this
galaxy. We detect both the [OII] and [OIII] lines and from these deduce an
average redshift of z = 0.034 (z=0.035 previously quoted by Tadhunter et
al. 1993). The best fit to the continuum of this galaxy is with a single
15 Gyr component. However, this fit is not ideal ($\chi^2$ = 1.33),
probably as a result of differential refraction, since the airmass for the
observations of this source was quite high (1.33).

\noindent {\bf 1514-24 (Ap Lib)} This is a BL~Lac object, classified on
the basis of its spectrum and light distribution (e.g. see Disney et al.
74). It is not possible to gain an adequate fit to the continuum with
either an old elliptical galaxy component or a young stellar component
alone or a combination of the two. Instead a power-law component is
clearly required and is the dominant component. We get reasonable fits
($\chi^2$ $\sim$ 0.6) with a 0.05-15 Gyr component combined with a
power-law component, with little difference between them. The [OIII] was
detected and from this we derived a redshift of z = 0.0480 .

\subsection{Far-Infrared/Starburst Link}

In addition to the UV excess, an alternative method for diagnosing the
presence of a starburst component in powerful radio galaxies is to use the
far-IR excess (e.g. Sanders \& Mirabel 1996).  However, the heating
mechanism for the dust radiating the far-IR emission is controversial:
heating by an AGN is a viable alternative to heating by a starburst. In
this context, it is interesting to consider whether there is any link
between the detection of a young stellar population and the detection of a
far-IR excess for the FRI sources in our sample. Recently we found
evidence for just such a link in samples of more powerful radio sources
(Tadhunter et al. 1996, Wills et al.  2002, Tadhunter et al. 2002).

In Table 2 we show which of the galaxies in our sample are detected 
in the IRAS `all-sky' survey (as described by Neugebauer   
et al. 1984) and/or  `pointed observations' of desired targets. The
pointed observations are the most sensitive observations made with
IRAS and for three of our sources such observations exist and have been
quoted from Golombek et al. (1988) and Impey \& Neugebauer (1988). 

The detections were estimated by co-adding scans from the all-sky
survey passing within approximately 1.7$^{\prime}$ of the target
position. A baseline was fit to each individual scan and the scans
were then co-added and analysed. An infrared source observed in a 60
$\mu$m co-added scan was presumed to be identified with a radio galaxy
if the positional agreement was better than 1$^{\prime}$, the peak
flux was greater than or close to three times the rms deviation of the
residuals after baseline subtraction, and the detection could be
confirmed at at least one other wavelength band (either 12 $\mu$m or
25 $\mu$m) and/or with pointed observations. In the case of a
detection, the flux density was deduced from the best-fitting point
source template of the co-added median scans. For the remaining
sources, in which no identification with an infrared source could be
made, upper limits were derived from three times the median rms
deviation of the residuals after baseline subtraction in a
20$^{\prime}$ region centred on the target position, outside of the
central 2.5$^{\prime}$ signal range. Flux calibration followed the
procedure outlined by Young et al. (1988) and conversion of the flux
density to a 60 $\mu$m luminosity was made assuming an infrared
spectral index of -1 (using flux density f$_{\nu}$ $\propto$
$\nu^{+\alpha}$, as used in Golombek et al. 1988). In Figure 3 we plot
the 60 $\mu$m luminosity of the sources of the sample against their
redshift on a log-log scale.

Using these criteria we find seven clear detections (0521-36, 0620-52,
0625-35, 0915-11, 1318-43, 1333-33, 1514-24) and 2 marginal detections
(0453-20, 1246-41).  In the case of 0453-20 the detection is only marginal
because of a relatively large positional offset between the 60$\mu$m and
optical galaxy positions ($\sim$1$^{\prime}$), while in the case of
1246-41 the detection is considered marginal because of relatively poor
S/N. Note that, based on the all-sky survey alone, the 60$\mu$m detection
of 0915-11 (Hydra A) would be considered marginal, but a flux has also
been measured for this source at 60 $\mu$m and 100 $\mu$m from pointed
observations (Golombek et al. 1988), albeit at a low level of accuracy.
The pointed and all-sky survey 60$\mu$m fluxes for this source agree
within 50\%. Therefore, on balance, we regard the detection of this source
as secure.

Although the three BL~Lac objects in our sample were detected by IRAS,
it is likely --- given the clear detection of a power-law component at
optical wavelengths --- that their far-IR fluxes are significantly
boosted by non-thermal synchrotron emission. Therefore, we exclude
these BL~Lacs from the following discussion on the optical
starburst/far-IR link.

Our results show that all three of the objects with the best evidence for
young stars are either detected or marginally detected by IRAS; the far-IR
luminosities of these sources are, in general, significantly larger than
those of the sources in which young stars were not detected (see Figure
3). It is also notable that, of the objects definitely not detected by
IRAS, none show evidence for young stellar populations. These results
point to a link between far-IR and optical starburst activity similar to
that found for samples of more powerful radio sources (Wills et al. 2002,
Tadhunter et al. 2002). However, more sensitive far-IR observations of the
low-luminosity sources will be required to put this link on a firmer
footing.

\begin{figure*}
\setlength{\unitlength}{1mm}
\label{fig3.fig}
\begin{picture}(10,80)
\put(0,0){\includegraphics{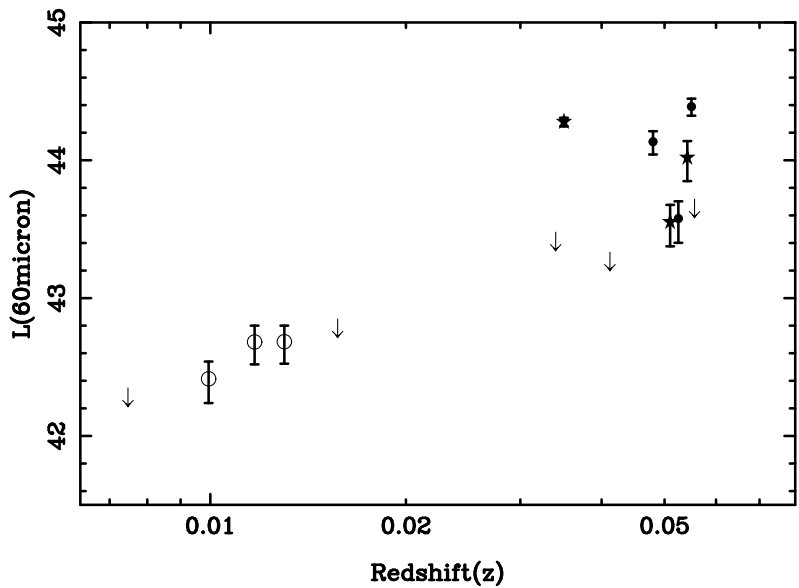}}
\end{picture}
\caption{The 60 $\mu$m luminosity of the sources of the sample plotted
against redshift presented as a log-log plot. The filled star symbols
represent those sources showing evidence of young stars and the filled
circles represent the BL~Lac objects. The sources which show no UV
excess but are detected by IRAS are represented by open circles and
those which show no UV excess and are also not detected by IRAS are
represented by arrows indicating the upper limit to the 60 $\mu$m
luminosity.}
\end{figure*}

\section{Discussion}

\subsection{UV excess in low power radio galaxies}

The presence of a UV excess in powerful radio galaxies compared with
quiescent ellipticals has been well known for some time. However, the
origin of this excess has been matter of discussion, although its
multi-component nature is now clear (Tadhunter et al. 2002). Although
the D(4000) parameter has previously been measured in some FRI sources
(Owen et al. 1996), this paper represents the first time that the UV
excess in low luminosity radio galaxies has been carefully
investigated in a complete sample, using both the D$^{\prime}$(4000)
parameter and the fit to the optical continuum.

Before analysing in detail the results for the low-luminosity sample,
it is worth summarising what has been found for powerful radio
galaxies (e.g. see Wills et al. 2002 and Tadhunter et al. 2002). For
the low redshift sample (Wills et al. 2002), three out of nine of the
galaxies show a UV excess which is attributed to a young stellar
population and none of these sources require a significant power-law
contribution to the excess. This has been found despite the higher
radio power and the strong emission lines characteristic of these
objects, in contrast to the FRIs studied here.  In the higher redshift
radio galaxies (Tadhunter et al. 2002), a UV excess is observed in all
of the sources of the sample, although the contribution to the UV
excess by a young stellar population is found to be a similar fraction
of objects ($\sim$ 30 - 50 \%) to that of the low redshift sample.

\subsubsection{Young stellar populations in low-luminosity radio galaxies}

From the D$^{\prime}$(4000) parameter we found that 5 (0453-20,
0620-52, 0625-35, 0915-11, 1514-24) out of the 13 observed galaxies
have a value indicating a possible additional blue component. However,
only by performing the fit to the continuum are we able to really
identify the nature of this component.  A young stellar population
component (0.05 - 2 Gyr) gives a substantial contribution to the UV
excess in three of these five galaxies (0453-20, 0620-52 and 0915-11).
For the remaining two, the dominant additional component is a
power-law. The latter is expected in the case of 1514-24 since this is
a well known BL~Lac object.  However, as discussed in the notes, the
second object, 0625-35, also appears to have the characteristics of a
BL~Lac.  This trend is also confirmed for the third BL~Lac in the
sample, 0521-36, for which a power-law component was found from data
in the literature.

Excluding the BL~Lacs, three out of eleven galaxies in our sample show a
significant contribution from a young stellar component. This fraction is
similar to that found for powerful FRII radio galaxies ($\sim$ 30\%; see
Wills et al. 2002 and Tadhunter et al. 2002). Thus, {\sl the fraction of
galaxies with a young stellar population component does not seem to depend
on the power of the galaxies}.  Assuming that the radio activity is
triggered by a merger, which also triggered a starburst, then this perhaps
indicates that there is no major difference in the origin, `triggering'
and time-scale of the activity in both kinds of radio sources.

\subsubsection{The power-law component} 

Another interesting result of this study, is that we only find
definite evidence for the presence of a power-law component in the
three BL~Lac objects.  From our data, therefore, it appears that {\sl
a power-law component is not required in the fit to the continuum of
the host galaxy of FRI radio sources}.

This result is intriguing since, at first sight, it appears to be in
disagreement with recent HST results.  As already described in \S 1,
optical compact cores have been found in a large fraction of galaxies
hosting FRI radio sources (e.g. Capetti \& Celotti 1999). The
correlation between their flux and the flux of the radio cores for
FRI/BL~Lacs indicates that these optical cores are due to optical
synchrotron radiation, supporting the similarities between FRI and
BL~Lacs as predicted by unified schemes. Given the low amount of
obscuration predicted (e.g. see Chiaberge et al. 1999, Morganti et
al. 2001) we would therefore also expect a power-law component in a
number of FRI radio galaxies. However, the cores detected by the HST
are actually quite weak (typically 10$^{-17}$ to 10$^{-16}$ erg
cm$^{-2}$ s$^{-1}$ {\AA}$^{-1}$) and therefore, they represent only a minor
contribution (i.e. only a few percent or less) to the optical flux for
a typical ground-based extraction aperture.  The typical extraction
aperture used here (2 $\times$ 7 arcsec), depending on the redshift of
the individual source, corresponds to approximately several
kiloparsecs.  On this scale the contribution of the host galaxy
represents a major fraction.

\subsection{Radio-optical correlation: is there a difference between FRIs 
and IIs?}

By subtracting the continuum from the spectra of the galaxies we have
detected emission lines in six galaxies where previously only upper
limits were available.  This allows us to further investigate the
correlation between radio power and emission line luminosity.  With
all the available data on the 2Jy sample we have constructed separate
correlations for 5 GHz radio power versus the luminosities of [OII]
and [OIII] emission lines respectively (see Figure 4).  In this way,
we can extend the work carried out by Tadhunter et al. (1998) to the
low radio power regime.

\begin{figure*}
\setlength{\unitlength}{1mm}
\label{fig4.fig}
\begin{picture}(10,90)
\put(0,0){\includegraphics{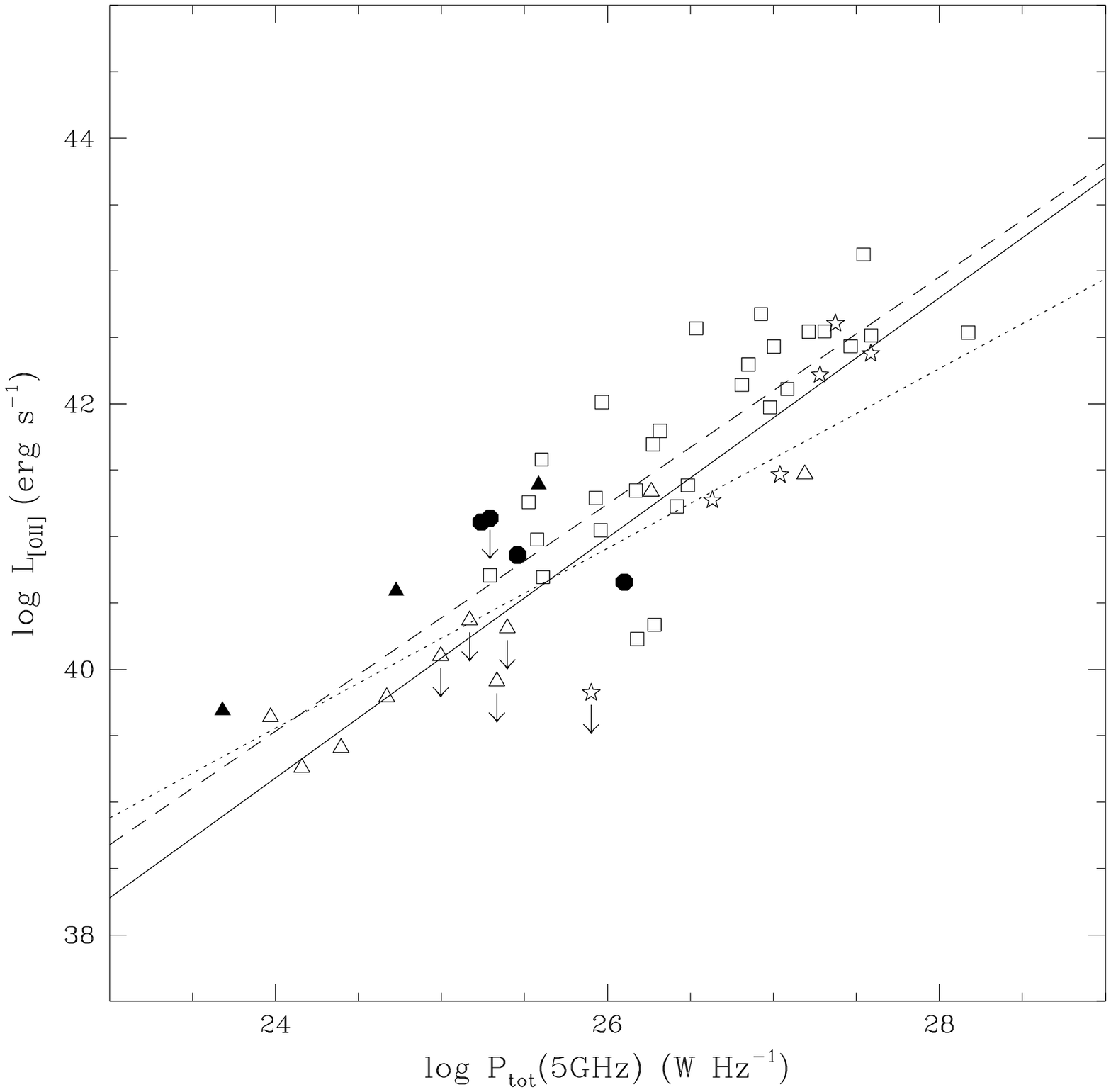}}
\put(0,0){\includegraphics{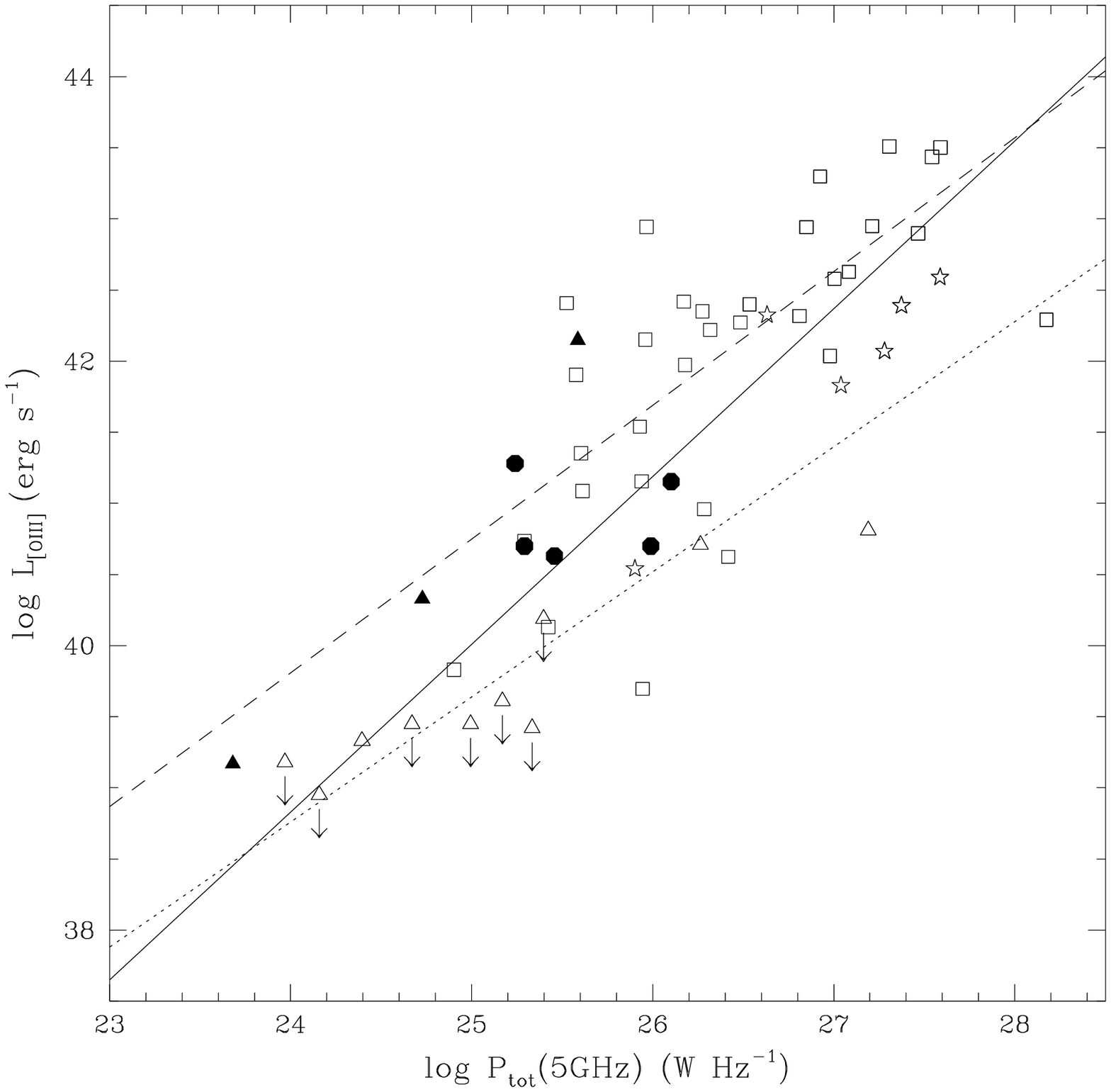}}
\end{picture}
\caption{ The correlation between radio power versus [OII] (left) and
[OIII] (right) emission line luminosities for the 2Jy sample of radio
galaxies (see text for further details). The objects are classified
according to their radio properties as follows: open squares - FRII;
open triangles - FRI; open stars - compact steep spectrum (CSS) radio
sources; filled circles - BL~Lac objects including two additional
objects (1807+69 (log P$_{tot}$ = 25.2, log L$_{[OII]}$ = 41.1, log
L$_{[OIII]}$ = 41.3) and 2200+42 (log P$_{tot}$ = 26.0, log
L$_{[OII]}$ not available, log L$_{[OIII]}$ = 40.7)) from Stickel,
Fried \& Kuehr (1993); filled triangles - uncertain radio
classifications. In both plots, the solid line represents the best fit
to all data in the sample, the short-dashed line represents the fit to
the FRI and BL~Lac objects and the long-dashed line is the fit to the
FRII and CSS objects. Details of the fitting procedure and resulting
fits are described in the text and in Table 4.}
\end{figure*}

In Figure 4, for the higher redshift range, we have taken all the
radio galaxies from the 2Jy sample described in Tadhunter et
al. (1998) with z $>$ 0.1 (a complete sample consisting of mainly FRII
sources but also some compact radio sources and FRI/FRII transition
sources, including 1648+05 (Her A)). For the lower redshift range, we
have included all the sources from our complete sample presented here
and all the radio galaxies with z $<$ 0.1 from Tadhunter et al. (1993:
this comprises a complete sample of FRIIs only since the FRIs/BL Lacs
with z $<$ 0.1 are already included in our sample). In this way, we
are actually comparing detections and upper limits resulting from
spectra of similar quality for both low-power and high-power radio
sources.

The analysis of the correlations has been carried out using the ASURV Rev
1.2 package (Lavalley, Isobe \& Feigelson 1992, Isobe, Feigelson \& Nelson
1986, Feigelson \& Nelson 1985) in order to take into account the upper
limits. In our analysis, the 5 GHz radio power was always considered as
the independent variable. The parameters obtained and the significance of
the correlations are summarized in Table 4 including the probability that
the two quantities are independent. Note that, although at 5 GHz the
BL Lac sources will probably have a large component of beamed flux, the
results of our correlations (within the errors) are not affected by the
removal of these sources from the data entirely and therefore this effect
is not significant.

\begin{table*}
\centering
\caption{Correlation analysis for the 2Jy sample shown in Figure
4. The analysis has been performed separately for the full sample (see
solid line in Figure 4), the FRI and BL~Lac objects (see short-dashed
line) and the FRII and CSS objects (see long-dashed line). The
parameters of the fits are presented and the final column gives the
significance level of the correlation (i.e. the percentage probability
that the correlations could arise by chance, see text for more
details).}
\label{tab4}
\begin{tabular}{ccccc}
\hline\hline\\
{\bf Correlation} & {\bf Sample}   &  \multicolumn{2}{c}{\bf Best fit}  & {\bf Prob.}  \\
                  &          & {\bf Slope} & {\bf Intercept Coeff.}  &    \%        \\
\hline    
log P$_{\rm 5GHz}$ {\sl vs} log L$_{\rm [OIII]}$ & All &  1.18 $\pm$ 0.11  &  10.51 $\pm$ 2.83  &  $<$0.01 \\
 & FRII + CSS &  0.94 $\pm$ 0.15  & 17.25 $\pm$ 3.86 & $<$0.01 \\
 & FRI + BL~Lac & 0.88 $\pm$ 0.27  &  17.64 $\pm$ 6.80 &  0.1 \\
log P$_{\rm 5GHz}$ {\sl vs} log L$_{\rm [OII]}$ & All & 0.91 $\pm$ 0.07  &  17.35 $\pm$ 1.91  & $<$0.01 \\
 & FRII + CSS & 0.86 $\pm$ 0.12  &  19.01 $\pm$ 3.20  & $<$ 0.01 \\
 & FRI + BL~Lac & 0.69 $\pm$ 0.14  & 22.93 $\pm$ 3.51   & 0.4  \\
\hline \\
\end{tabular}
\end{table*}

For all available data on the 2Jy sample, we find that there is a
clear correlation between the radio power and both the [OIII] and
[OII] line luminosities. The presence of a correlation between line
luminosity and radio power is already well known from a number of
studies (e.g. see Baum \& Heckman 1989a,b, McCarthy 1988, Morganti,
Ulrich \& Tadhunter 1992, Rawlings et al. 1989, Tadhunter et
al. 1998). However, in the previous studies, the correlation was
mainly investigated using the luminosity of only {\sl one} emission
line (usually either [OIII] or H$\alpha$).  In the single previous
case where two lines were used ([OIII] and [OII], Tadhunter et
al. 1998) the sample was limited to powerful radio galaxies (z $>$
0.1).  Our data, therefore, allows a significant expansion on previous
results.

From the parameters of the correlations listed in Table 4, it appears
that [OIII] and [OII] follow correlations with two marginally
different slopes, with the latter showing a somewhat flatter
slope. The two correlations also have a different scatter: the
correlation with [OII] has a much smaller scatter than that of the
[OIII]. This was already noted for the sample of mostly powerful radio
galaxies studied by Tadhunter et al. (1998). The explanation of this
result is provided by the photoionization model, where the [OIII]
lines are more sensitive to variation in the ionising continuum.

It is interesting to investigate from the new data whether the
correlations of the radio power versus emission line luminosity are
different for low-power FRI and high-power FRII galaxies. This is an
important question since it may point to different mechanisms for
ionising the gas in the two groups of galaxies (see Baum et al. 1995).
When we split the sample into FRI and FRII radio morphologies, we find
that (see Figure 4 and Table 4), {\sl FRIs follow the correlation
between line luminosity and radio power with a similar slope to that
found for the FRIIs}.  The significance of the correlations (in
particular between [OII] luminosity and radio power) is, however,
lower for the FRIs than for the FRIIs.

The slope found in the [OII] correlation is consistent (within the errors)
with the result obtained by Morganti et al. (1992) using H$\alpha$ lines
for a sample of low-luminosity radio galaxies taken from the B2 sample and
a sample of 3C powerful radio galaxies (from Baum \& Heckman 1989a,b). It
is, however, much steeper than the slope derived by Zirbel \& Baum (1995)
for a large but heterogeneous sample of objects collected from the
literature (mainly using H$\alpha$+[NII] emission line luminosities). It
is likely that this difference is due to the uncertainties involved in
their fitting of a combination of samples selected using different
criteria which includes the mixing of emission lines ([OIII] is used when
H$\alpha$ is unavailable), the mixing of radio frequencies (5GHz
measurements are used when no good 408 MHz observations are available)  
and, most importantly, the inclusion of a number of sources where no
emission line luminosity information is available. This latter issue has
the effect of introducing an emission line flux density limit since the
fainter sources (mostly FRIs) are more likely to go undetected. Since the
sample is already radio flux density limited, this results in difficulty
interpreting any correlation observed between radio flux and emission line
luminosity, particularly for the FRIs. In contrast, our sample, the
Morganti et al. sample and the Baum \& Heckman samples mentioned above are
all based on homogeneous data and result in consistent measurements of the
slope. 

Despite the similarities mentioned above, an offset is evident in the
[OIII] luminosity versus radio power plot, between the correlation for
FRIs and FRIIs.  This can be seen even more clearly in the histogram
of Figure 5 showing the ratio between the optical luminosity and the
radio power. Here we compare all the FRIs from our sample with all
FRIIs from Tadhunter et al. (1993) within the same redshift range
(i.e. z $<$ 0.06). The difference is highly significant (with a
probability of 3\% that the two distributions are similar,
estimated using ASURV).  The observed offset confirms the results
obtained by Zirbel \& Baum (1995) and Baum et al. (1995) since the FRI
galaxies, for a given radio power, have systematically lower optical
line luminosity than their powerful cousins.

\begin{figure}
\setlength{\unitlength}{1mm}
\label{fig5.fig}
\begin{picture}(10,85)
\put(0,0){\includegraphics{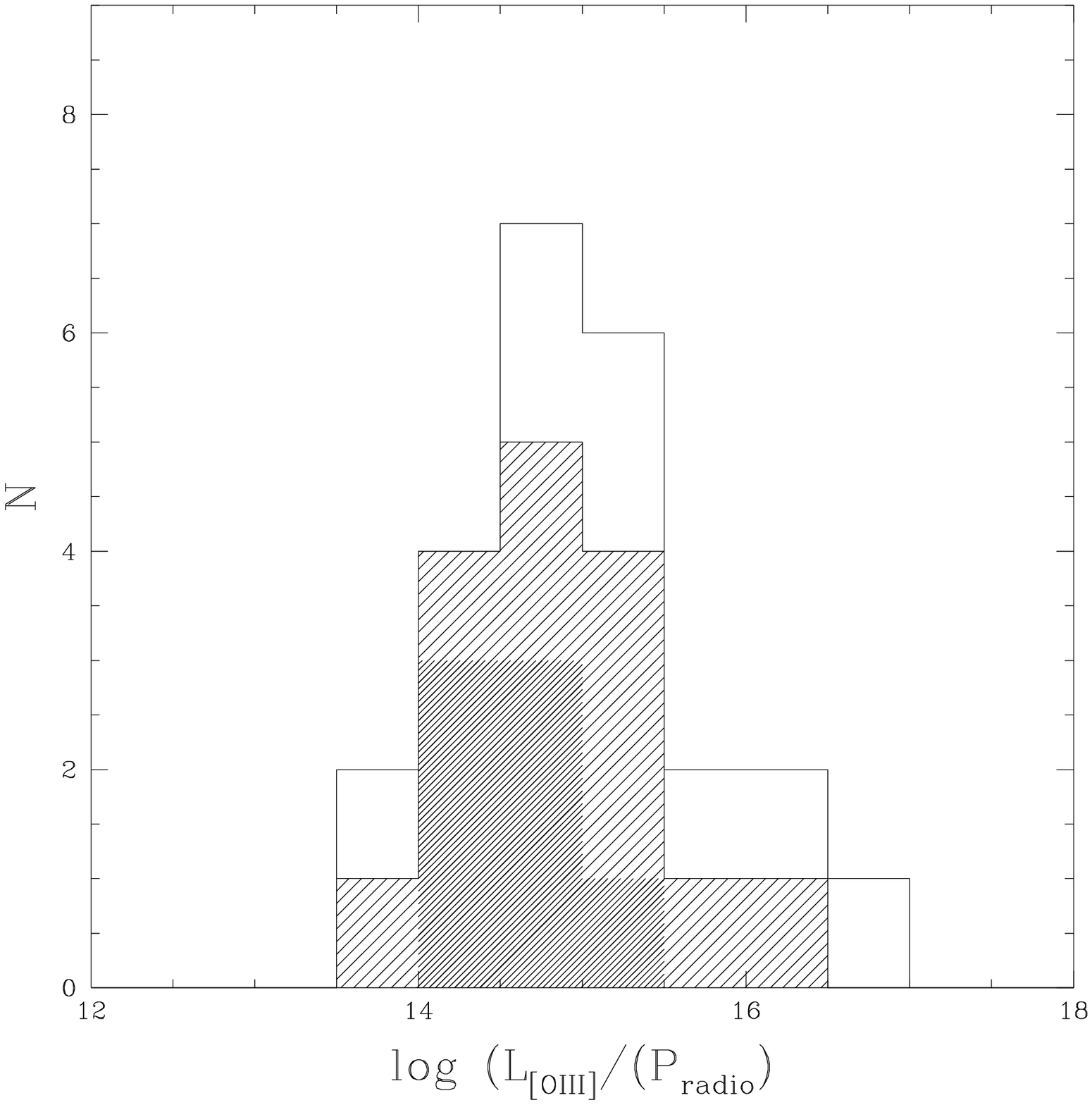}}
\end{picture}
\caption{Distribution of the ratio L$_{[OIII]}$/P$_{radio}$ for the
2Jy sample. FRIs are shown as large dashed and the upper limits for
the FRIs are shown as narrow dashed; the rest of the histogram
represents the FRIIs. See text for more details.}
\end{figure} 

In summary, we find that the low-luminosity FRI radio galaxies follow
a correlation between the luminosity of the lines (both in [OII] and
in [OIII]) and their radio power, and that the slope of these
correlations is not too dissimilar from the one found for FRIIs. This
result supports the idea that in FRIs, as in FRIIs, nuclear UV
radiation is responsible for the ionization of the gas around the AGN.
This conclusion is quite different from that reached by Baum et
al. (1995): their work suggests that two different mechanisms are
responsible for the ionization of the lines, on the basis of their
finding of a large difference in slope for FRIs compared with FRIIs.
However, it is indeed the case that FRIs have a systematically lower
line luminosity (for their radio power) compared with FRIIs, although
this is mainly the case for the [OIII] luminosity. Interestingly, at
the transition between FRI and FRII (around 10$^{25.5}$ at 5GHz) the
[OIII] luminosities seem to become suddenly stronger and show a `step'
in their distribution, a feature which is not detected in [OII] (see
Figure 4).  A possible explanation of this is that, based on the
photoionization models, [OIII] is much more sensitive to changes in
the ionising continuum than [OII]. Below a critical ionization
parameter of U $\sim$ 10$^{-3}$ the photoionization models show that
the correlation between the [OIII] luminosity and the ionization
continuum steepens (from 1.2 to 3.3). Thus, if the ionising continuum
luminosity were to be significantly less for the FRIs than the FRIIs,
this would be much more apparent in [OIII] than [OII] - perhaps
leading to the `step' seen in the [OIII] plot.

Thus, the differences between FRIs and FRIIs could reflect a change to
a different mode of accretion: advective low efficiency flow (e.g. see
Rees et al. 1982) for FRIs and standard optically thick accretion
disks (e.g. see V$\acute{e}$ron \& V$\acute{e}$ron-Cetty 2000) for
FRIIs. Since the latter would be expected to produce ionising photons
more efficiently, this is qualitatively consistent with what we
observe, and may also be consistent with the failure to detect broad
lines in FRIs (e.g Capetti et al. 2002).

\subsection{FRI, BL~Lacs and unified schemes} 

According to the unified schemes for radio loud objects, FRIs are
believed to be the parent population of BL~Lac objects (Urry \&
Padovani 1995).  If this is the case, we would expect the luminosity
of the emission lines in the two groups of objects to be similar.
This is because the emission lines are not expected to suffer from any
beaming effect.

This important test for the unified schemes has already been carried
out for FRIIs and radio-loud quasars. In this case, differences were
found in the luminosity of the [OIII] lines (with quasars showing
emission line luminosities typically 5-10 times higher than, otherwise
similar, FRIIs; Jackson \& Browne 1990) while the luminosities of the
[OII] lines were found to be similar in the two groups (Hes, Barthel
\& Fosbury 1996).  This result has been explained in terms of
obscuration from the circumnuclear torus which affects high ionization
lines more than the low ionization lines since the former are believed
to originate closer to the nucleus.

In the case of the FRIs and BL~Lacs, if we are to assume that no
significant absorption occurs in the FRI sources, then we should
expect no difference between the [OII] and [OIII] emission line
luminosities. However, the question of whether obscuration exists in
FRIs is still a matter of discussion (Chiaberge et al. 1999, Morganti
et al. 2001, Chiaberge et al. 2002, Hardcastle et al. 2002). For
example, recent work by Chiaberge et al. (2002) has suggested that
there is clear evidence of nuclear absorption in some FRI sources but
not by a `standard' torus-like structure. Instead they suggest that
the moderate amount of absorption observed might be accounted for
either by extended kpc scale dust lanes or by $\sim$ 100 pc scale
dusty disks.

Urry \& Padovani (1995) have already attempted a comparison between these
two groups, using data available in the literature. Interestingly, their
results tentatively suggest an apparent difference between the [OIII]
luminosities of BL~Lacs and FRIs, with the former stronger than the latter.
However, their comparison uses data from different samples and the FRI
fluxes may well be underestimated due to small slits and contamination from
a strong stellar continuum. To determine if this [OIII] difference is
really significant, we now repeat this study using our complete sample,
which consists of high quality optical spectra for which the continua have
been accurately subtracted.  Using ASURV, we have estimated both the mean
of the distributions as well as the probability that the two distributions
are drawn from the same parent population. The statistical tests used
within ASURV to compare the distributions are Gehan's generalised Wilcoxon
test, the Logrank test, the Peto-Peto test and the Peto-Prentice test.
These tests are generalisations of standard techniques based on ranking the
data values and comparing the distributions of the ranks. Consistent
results were obtained for all the tests which support the tentative
suggestion by Urry \& Padovani (1995) that the [OIII] luminosity of the
BL~Lacs is higher than the FRIs.  Since our sample only contains three
BL~Lacs, we have also repeated these tests including two additional objects
(1807+69 and 2200+42) from the compilation of Stickel et al. (1993), which
contains sources of a similar range in radio power to our sources. These
additional two sources are the only two from the sample of Stickel et al.,
together with ApLib, to have z $<$ 0.150 and line luminosity data
available. We find that similar results and conclusions can be drawn with
the addition of these two objects and the distributions of luminosity of
the emission lines and radio power for the FRI and BL~Lac sources
(inlcuding the Stickel sources) are presented in Figure 6. In Table 5 we
summarize the results of the statistical tests performed within ASURV. Note
that this gives the typical values obtained for the different tests used
within ASURV (inlcuding the Stickel sources) and is in good agreement with
the results obtained without the addition of the Stickel objects.

\begin{table} 
\caption{Comparison of the distributions of the BL~Lac
sources and FRI sources for the 14 radio galaxies in our sample plus two
additional BL~Lac objects (1807+69 and 2200+42) taken from Stickel et al.
(1993). The final column gives the percentage probability that the two
distributions are drawn from the same parent population. See text for more
details.}
\label{tab5} 
\begin{tabular}{ccccc} \hline\hline\\ 
{\bf Sample} & {\bf
Mean} & {\bf Probability (\%)} \\
\hline
{\bf [OII]} BL~Lacs    & 40.820 $\pm$ 0.096  &  \\
       FRI     & 39.960 $\pm$ 0.235  &  $\sim$ 6    \\ 
\hline
{\bf [OIII]} BL~Lacs    & 40.840 $\pm$ 0.156  &     \\ 
       FRI     & 39.509 $\pm$ 0.213  &  $\sim$ 0.8      \\ 
\hline
 {\bf Radio} BL~Lacs & 25.260 $\pm$ 0.236  &       \\
       FRI     & 25.220 $\pm$ 0.322  &  70         \\ 
\end{tabular}
 \end{table}

From these results it appears that {\sl while the difference between
FRIs and BL~Lacs is marginal for the [OII] line luminosities, there is
a significant difference in the luminosity of the [OIII] lines}: The
[OIII] luminosity of the BL Lacs is higher (40.840 pm 0.156) than the
FRIs (39.509 pm 0.213). This is an interesting result in view of the
current debate on obscuration in FRIs, since this does indeed suggest
that a significant amount of obscuration exists. Of course our result
should be taken with some care, since it has been derived from a small
sample. Nevertheless, it is in agreement with the tentative suggestion
made by Urry \& Padovani (1995).

An alternative explanation to obscuration that must be investigated is the
possibility that some selection effect has been introduced in the BL~Lac
sample. We therefore looked at the distribution of total radio power for
FRI and BL~Lacs (see Figure 6 and Table 5) but no difference was found. In
fact, since the radio powers of three of the BL~Lacs are dominated by
the core emission, this suggests that the BL~Lacs in our sample have much
lower extended radio powers than the total powers displayed in Figure 4.
Therefore, the BL~Lac sources should be compared with the FRI sources in
the sample at the lower end of the emission line luminosity and radio power
distribution. This strengthens our finding that the BL~Lacs have higher
[OIII] luminosities than the FRIs in the sample.

Another possibility is that {\sl FRIs do not represent the parent
population of all BL~Lacs}. The suggestion that the simple picture of
the unified scheme is probably not adequate has already been made
(Kollgaard et al. 1992, Murphy, Browne \& Perley 1993) and it has also
been put forward that weak-lined FRIIs could be associated with
BL~Lacs (and not with quasars; Jackson \& Wall 1999).

\begin{figure*}
\setlength{\unitlength}{1mm}
\label{fig6.fig}
\begin{picture}(10,120)
\put(0,0){\includegraphics{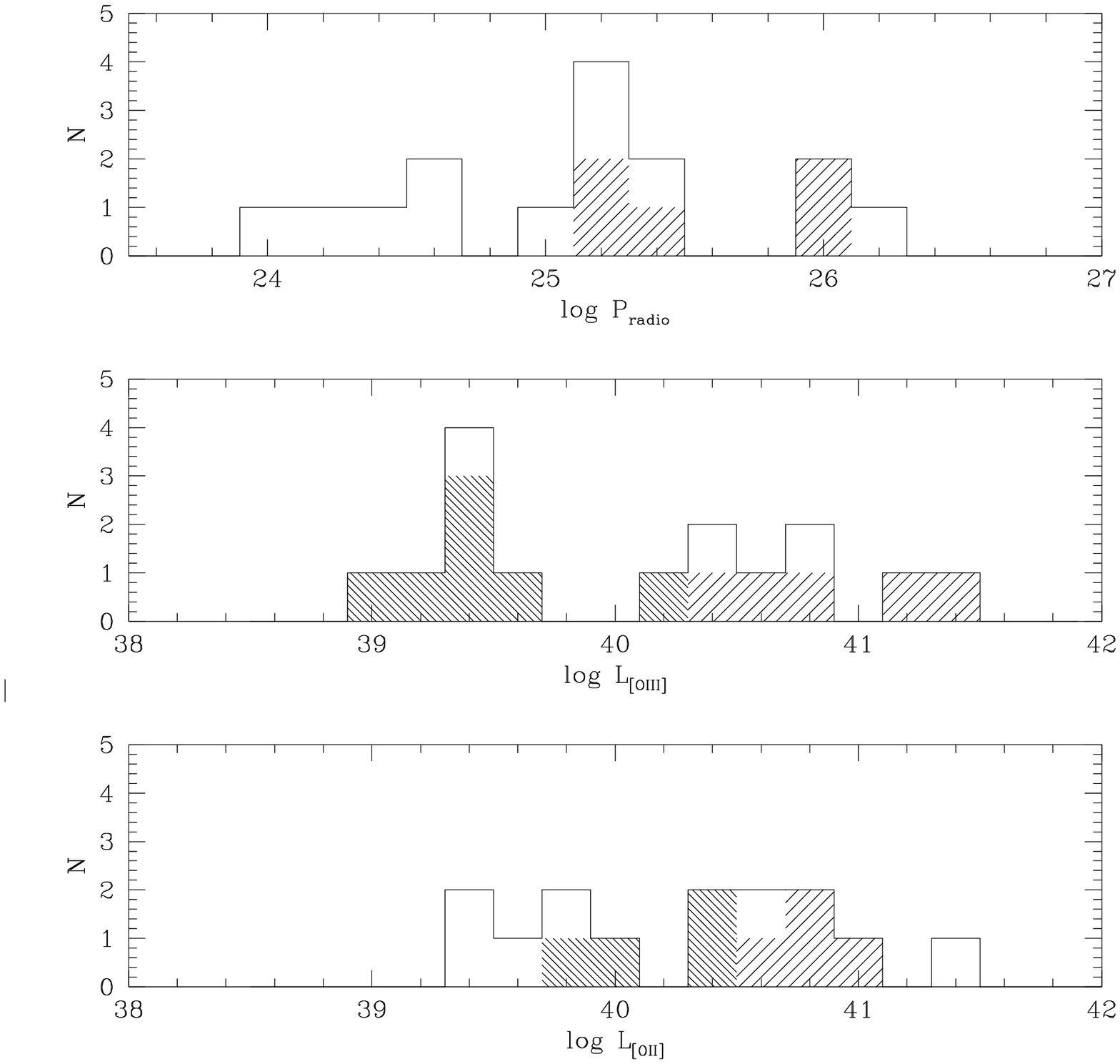}}
\end{picture}
\caption{Distribution of total radio power, [OIII] and [OII] line
luminosities for the FRI and BL~Lac sources in our sample, plus two
additional BL~Lac objects (1807+69 and 2200+42) taken from Stickel et
al. (1993). The BL~Lacs are shown as large dashed and the upper limits
for the FRIs are shown as narrow dashed; the rest of the histogram
represents the remaining FRIs.}
\end{figure*}

\section{Conclusions}
 
In a complete sub-sample of 13 low-luminosity radio galaxies, we find
that 5 show a UV excess and, in 3 cases, this excess appears to be due
to the presence of young stars. The remaining two sources which show a
UV excess are BL~Lac objects. Excluding the BL~Lac objects, we
therefore find that $\sim$ 30\% of our low-luminosity sample show
evidence for a young stellar population. Since this is similar to the
fraction found for powerful FRII radio galaxies, we suggest that the
proportion of galaxies with young stars does not depend on the power
of the galaxies. Furthermore, our modelling also suggests that a
power-law component is not required in the fit to the continuum of the
host galaxy of FRI radio sources. 

We find that the three objects which show evidence for young stars are
either detected or marginally detected by IRAS and are among the
highest 60 $\mu$m luminosity sources in the sample. Of the objects
definitely not detected by IRAS, none show evidence for young stellar
populations. This suggests a link between far-IR and optical starburst
activity similar to that found for samples of more powerful radio
sources.

On studying the correlations between radio power and the [OIII] and
[OII] line luminosities of the 2Jy sample we find that the FRI sources
follow the correlations with a similar slope to that found for the
FRIIs. This result supports the idea that in FRIs, as in FRIIs,
nuclear UV radiation is responsible for the ionization of the gas
around the AGN.

Our investigation into the luminosity of the emission lines in the FRI
and BL~Lac sources has shown that there is a significant difference in
the [OIII] line luminosities of the two groups, whilst the difference
in [OII] is marginal. A similar result has also been found for FRIIs
and radio-loud quasars, which has been explained in terms of
obscuration of the high ionization lines which originate closer to the
nucleus. To explain the same result in our low-luminosity sample we
must either invoke some kind of obscuration model or conclude that the
two groups cannot actually be unified.

\subsection*{Acknowledgments} 

We acknowledge the expert advice and assistance of the scientific
support staff and the data analysis team at the Infrared Processing
and Analysis Facility (IPAC). We also acknowledge the data reduction
package FIGARO provided by the Starlink Project which is run by CCLRC
on behalf of PPARC. This research has made use of the NASA/IPAC
Extragalactic Database (NED) which is operated by the Jet Propulsion
Laboratory, California Institute of Technology, under contract with
the National Aeronautics and Space Administration.  KAW is supported
by a Dorothy Hodgkin Royal Society Fellowship.

{}

\end{document}